\documentclass[10pt,a4paper]{article}
\usepackage{moreverb, bm}
\usepackage[colorlinks,bookmarksopen,bookmarksnumbered,citecolor=red,urlcolor=red]{hyperref}
\usepackage{graphicx}

\pdfoutput=1

\usepackage[mathscr]{euscript}
\usepackage{amsmath}

\newcommand{\rmd}{{\rm d}}
\newcommand{\rme}{{\rm e}}

\newcommand{\erf}{\textrm{erf}}

\newcommand{\be}{\begin{equation}}
\newcommand{\ee}{\end{equation}}
\newcommand{\bd}{\begin{displaymath}}
\newcommand{\ed}{\end{displaymath}}

\renewcommand{\be}{\ensuremath{\mathbf{e}}}

\newcommand{\bz}{\ensuremath{\mathbf{z}}}

\newcommand{\bbeta}{{\mbox{\boldmath $\beta$}}}

\newcommand{\btheta}{{\mbox{\boldmath $\theta$}}}

\newcommand{\Prob}{\mathscr{P}}

\newcommand{\LL}{\mathscr{L}}

\newcommand{\dt}{{\rm d}t}
\newcommand{\ds}{{\rm d}s}
\newcommand{\dw}{{\rm d}w}

\DeclareMathOperator*{\argmax}{arg\,max}

\newcommand\BibTeX{{\rmfamily B\kern-.05em \textsc{i\kern-.025em b}\kern-.08em
T\kern-.1667em\lower.7ex\hbox{E}\kern-.125emX}}

\begin{document}

\title{Practical survival analysis tools for heterogeneous cohorts and informative censoring}

\author{M. Rowley\textsuperscript{a}, 
H. Garmo\textsuperscript{b},
M. van~Hemelrijck\textsuperscript{b},
W. Wulaningsih\textsuperscript{b},\\
B. Grundmark\textsuperscript{c,d},
B. Zethelius\textsuperscript{e,d},
N. Hammar\textsuperscript{f,g},
G. Walldius\textsuperscript{h},\\
M. Inoue\textsuperscript{i},
L. Holmberg\textsuperscript{b} and 
A.C.C. Coolen\textsuperscript{a}
}

\maketitle

\begin{abstract}
{\bf{In heterogeneous cohorts and those where censoring by non-primary risks is informative many conventional survival analysis methods are not applicable; the proportional hazards assumption is usually violated at population level and the observed crude hazard rates are no longer estimators of what they would have been in the absence of other risks. In this paper, we develop a fully Bayesian survival analysis to determine the probabilistically optimal description of a heterogeneous cohort and we propose a novel means of recovering hazard rates and survival functions `decontaminated' of the effects of any competing risks. 
Most competing risks studies implicitly assume that risk correlations are induced by cohort or disease heterogeneity that is not captured by covariates. We additionally assume that proportional hazards hold at the level of individuals, for all risks, leading to a generic statistical description that allows us to decontaminate the effects of informative censoring, and from which Cox regression, frailty and random effects models, and latent class models can all be recovered as special cases.  
Synthetic data confirm that our approach can map a cohort's substructure,  and remove heterogeneity-induced false protectivity and false aetiology effects. Application to survival data from the ULSAM cohort leads to plausible alternative explanations for previous counter-intuitive inferences to prostate cancer. The importance of managing cardiovascular disease as a comorbidity in women diagnosed with breast cancer is suggested on application to survival data from the AMORIS study.}}\\[5mm]
\footnotesize{\textsuperscript{a} 
Institute for Mathematical and Molecular Biomedicine, King's College London, London, U.K.}
\\
\footnotesize{\textsuperscript{b} 
Cancer Epidemiology Group, King's College London, School of Medicine, Guy's Hospital, London,  U.K.}
\\
\footnotesize{\textsuperscript{c}
Department of Surgical Sciences, Uppsala University, Uppsala, Sweden.}
\\
\footnotesize{\textsuperscript{d}
Medical Products Agency, Uppsala, Sweden.}
\\
\footnotesize{\textsuperscript{e}
Department of Public Health and Caring Sciences/Geriatrics, Uppsala University, Sweden.}
\\
\footnotesize{\textsuperscript{f}
Department of Epidemiology, Institute of Environmental Medicine, Karolinska Institutet, Sweden.}
\\
\footnotesize{\textsuperscript{g}
AstraZeneca Sverige, S\"odertalje, Sweden.}
\\
\footnotesize{\textsuperscript{h}
Department of Cardiovascular Epidemiology, Institute of Environmental Medicine, Karolinska Institutet, Sweden.}
\\
\footnotesize{\textsuperscript{i} 
Department of Electrical Engineering and Bioscience, Waseda University, Tokyo, Japan.}
\\[3mm]
\footnotesize{Mark Rowley, Institute for Mathematical and Molecular Biomedicine, King's College London,}\\
\hspace*{26mm} \footnotesize{Hodgkin Building,
London SE1 1UL, U.K. ~~~E-mail: mark.rowley@kcl.ac.uk}\\[3mm]
\footnotesize{\emph{Contract/grant sponsor: Prostate Cancer UK, European Union FP-7 Programme (IMAGINT), and the Ana Leaf Foundation}}
\end{abstract}

\clearpage
\section{Introduction}

The analysis of survival data is often complicated by cohort heterogeneity and informative censoring arising from competing risks. In this paper, we extend the conventional approaches to modelling a cohort and present a survival analysis which determines probabilistically the most likely characterisation of a cohort and can provide cause-specific hazard rates and survival curves `decontaminated' of the effects of informative censoring.

A cohort is subject to informative censoring if the event-times of the primary and non-primary risks are not statistically independent; a cohort for which this is true is described as being subject to `competing risks'. Unfortunately, one cannot infer presence or absence of risk correlations from survival data alone \cite{Tsiatis,Gail}, and in many cases the independence assumption which underpins many survival analysis approaches is expected to be incorrect. Residual heterogeneity, which is not visible in the covariates, is often a fingerprint of such risk event-time correlations in cohorts having competing risks. The inference of risk characteristics from survival data in the hypothetical situation where all other risks were disabled requires that the competing risk problem be addressed \cite{Klein}, that is, it is necessary to \emph{somehow} handle contamination by informative censoring. In discussing informative censoring in a cohort, we shall refer to the \emph{crude} cause-specific hazard rates and survival functions, in which the influence of competing risks is present, and we define their \emph{decontaminated} analogues as the cause-specific decontaminated hazard rates and survival functions, isolated from the effects of all other risks.

The effective determination of heterogeneity can lead to a more accurate understanding of the characteristics of a cohort. In clinical epidemiology, diagnosing and disentangling cohort heterogeneity is crucial in many studies where differences in ``case-mix" are considered to be problematic, e.g. in comparing treatment outcomes from different institutions. Unaccounted for risk correlations can lead to incorrect inferences \cite{AndersenReview,DiSerio,ScharfsteinRobins,Dignam,Thompson} and the importance of having reliable epidemiological tools for isolating statistical features even for interrelated comorbid diseases is increasingly recognised \cite{Soneji}. An understanding of how a modifiable covariate (e.g. taking a particular medication, consuming a particular food type, smoking, exercising, etc.) having different influences for different groups within a cohort could prove advantageous for effective monitoring and management of population health. In populations where longevity is increasing, once relatively uncommon conditions are becoming more prevalent with the development of more effective therapies to treat competing conditions. Knowledge of the expected survival against such currently uncommon diseases, decontaminated of the influence of other risks, may provide a useful means of allocating research and public health resources efficiently.

\begin{figure}[t]
\centering
\includegraphics{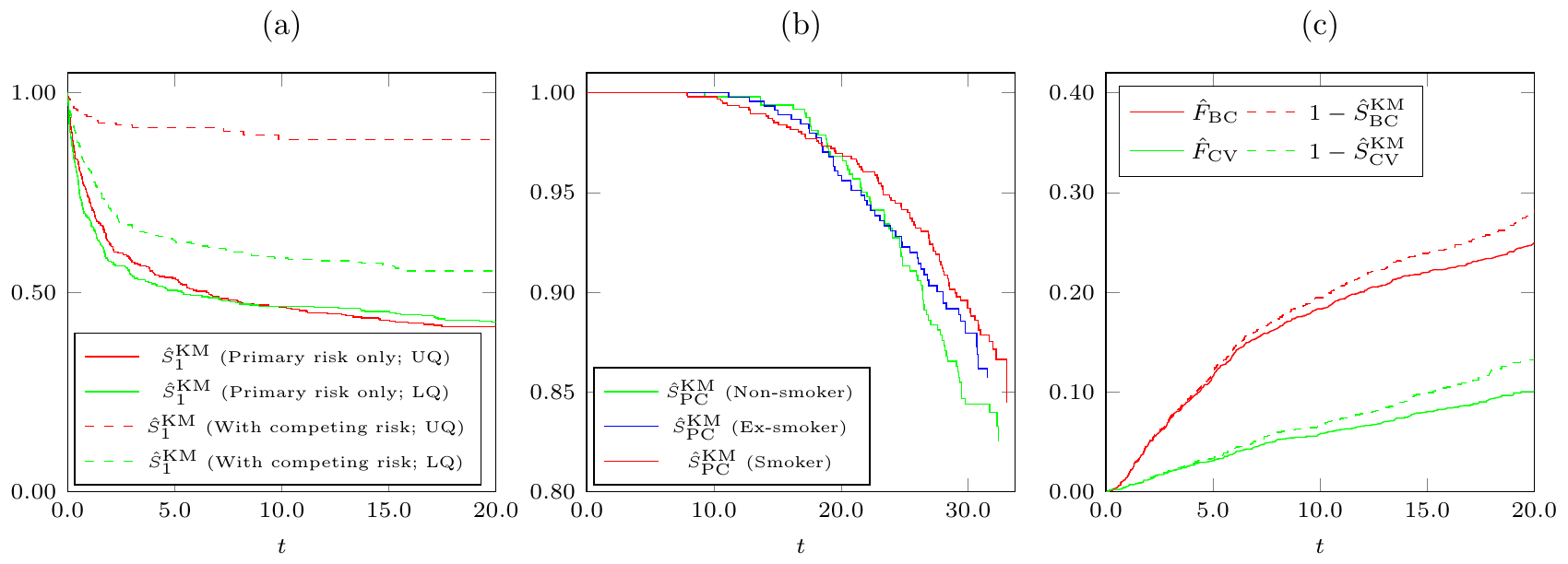}
\caption{\footnotesize{\emph{Effects and signatures of heterogeneity and informative censoring:} In (a) an illustration of the dangers of using covariate-conditioned risk-specific Kaplan-Meier (KM) estimators, $\hat{S}_r^{\rm{KM}}$, in the presence of competing risks. The estimators for the lower and upper quartiles (LQ and UQ respectively) of a covariate associated with the primary risk are shown for two simulated cohorts, one in which there is no informative censoring and another which is subject to a competing risk, with both cohorts sharing {\em identical} primary risk characteristics. These estimators offer an extremely misleading indication of survival against the primary risk in the presence of informative censoring (i.e. a competing risk), as can be seen by the differences between the estimators in the absence (solid lines) and presence (dashed lines) of informative censoring. The application of our practical tools to provide \emph{decontaminated} survival estimators, free from the effects of the competing risk, to the synthetic cohort data shown here is detailed in Section~\ref{sec:SynthData}. In (b) the use of the covariate-conditioned risk-specific Kaplan-Meier estimator to detect the violation of proportional hazards assumption and infer possible heterogeneity in a cohort. The Kaplan-Meier estimator for prostate cancer ($r=$PC) risk, conditioned on an individuals smoking status, is suggestive of heterogeneity in the ULSAM cohort; the estimator would suggest that the survival of smokers is greater than that of ex- and non-smokers (an alternative view is suggested from our analysis; Section~\ref{sec:ULSAM}). In (c) the empirically measured cumulative incidence, $\hat{F}_r(t)$, and the complement of the Kaplan-Meier estimator, $\hat{S}_r^{\rm{KM}}$, for breast cancer ($r=$BC) and cardiovascular disease ($r=$CV) death for women diagnosed with breast cancer from the AMORIS population; the difference between the cumulative incidence and the complement of the corresponding Kaplan-Meier estimator indicating the presence of competing risks in the cohort. }
}
\label{fig:synthetic_illustration}
\end{figure}

Simple survival analysis methods, such as Kaplan-Meier estimators \cite{KM} and Cox regression \cite{Cox}, are unable to correctly capture the characteristics of a cohort in the presence of heterogeneity as their successful use requires proportional hazards across the cohort as a whole. The uncritical use of these techniques can be quite misleading when the proportional hazards assumption is not satisfied, as demonstrated in Figure \ref{fig:synthetic_illustration}~(a) for a simulated cohort modelling false protectivity effects. However, they can offer clues as to the presence of heterogeneity and informative censoring in a cohort. Inspection of the covariate-conditioned risk-specific Kaplan-Meier estimators can confirm violation of the proportional hazards assumption, as is clearly evident for the cohort in Figure \ref{fig:synthetic_illustration}~(b). Differences between the cumulative incidence \cite{Kalbfleisch} and the complement of the corresponding risk-specific Kaplan-Meier estimator are indicative of competing risks in a cohort e.g. \cite{CompRisks_PracticalPerspective}, as shown in Figure \ref{fig:synthetic_illustration}~(c).

Many authors have tried to model residual cohort heterogeneity, usually starting from Cox-type cause-specific hazard rates, but with additional individualised risk multipliers. If the multipliers do not depend on the covariates we speak of `frailty models', e.g. \cite{Lancaster,Vaupel,Zahl,Yashin,Gorfine}, and regard them as representing the impact of unobserved covariates, see e.g. \cite{KeidingAndersenKlein}. If they depend on the covariates  we would speak of 
`random effects models', e.g. \cite{Vaida,DiSerio,Rosner,Wienke,Duchateau}. If the distribution of frailty factors takes the form of discrete clusters (latent classes, \cite{Lazarsfeld}), we obtain the latent class models; see e.g. \cite{HuangWolfe} or \cite{Muhten} (which combines frailty and random effects with covariate-dependent class allocation as in \cite{Reboussin}). Further variations include time-dependent frailty factors, and models in which the latent class 
of each individual is known. Most frailty and random effects studies, however, quantify only the hazard rate of the primary risk. They thereby capture some consequences of cohort heterogeneity, but without modelling also the non-primary risks it is fundamentally impossible to deal with the competing risk problem.

The approach of \cite{FineGray} focuses on parametrising the covariate-conditioned cumulative incidence function of the primary risk. It is conceptually similar to \cite{Cox}; both model the primary risk profile in the presence of all risks. Cumulative incidence functions appear more intuitive than hazard rates; they are directly measurable, and incorporate also the impact of non-primary risks. However, expressing the data likelihood in terms of cumulative incidence functions is more cumbersome than in terms of hazard rates. But while \cite{FineGray} quantify risks that compete, they do not address the competing risk problem.  Further developments  involve e.g. alternative parametrisations \cite{Fine,KleinAndersen}, application to the cumulative incidence of non-primary risks \cite{JeongFine}, and the inclusion of frailty factors \cite{Katsahian}.

Another community of authors have focused further on identifying which mathematical constraints or conditions need to be imposed on multi-risk survival analysis models in order to circumvent the identifiability problem, and infer the joint event time distribution unambiguously from survival data. Examples involving survival data with covariates are \cite{HeckmanHonore}, and \cite{AbbringBerg}. However, also these studies do not take the step towards decontamination tools or facilitate interpretation.

In our analysis, we build upon the frailty, random effects, and latent class approaches, to develop a generic model from which we estimate the relative frailty, covariate association(s), and base hazard rate for \emph{each} latent class and for \emph{all} risks. Additionally, by assuming that informative censoring at the cohort level is a consequence of residual (disease- or patient-) heterogeneity that is not captured by covariates in populations where only at the level of individuals are the different risks independent, we are able to derive exact formulae for decontaminated hazard rates and cause-specific survival functions. Our assumption of such heterogeneity-induced informative censoring is much weaker than assuming risk independence, yet it still imposes sufficient constraints to decontaminate from the effects of informative censoring. Furthermore, our analysis offers improvement over existing approaches in several distinct ways: 
In choosing to model \emph{all} risks \emph{simultaneously} more information can be extracted from cohort data in which risk correlations are present; in such circumstances, the application of an analysis approach which models the primary risk only is akin to choosing to not fully utilise the information available from the cohort data. Our analysis also introduces a fully Bayesian model selection for the determination of the optimal characterisation of a cohort.

In Section \ref{sec:HICR} we classify the distinct levels of ``risk complexity" in a cohort from the competing risk perspective and define precisely what we mean by heterogeneity-induced informative censoring before presenting the mathematical development behind our approach to survival analysis.

In Section~\ref{sec:SynthData} the effectiveness of our analysis to characterise heterogeneous cohorts is demonstrated; the results obtained on application of our analysis to synthetic survival data, simulating a variety of heterogeneous cohorts and informative censoring, are presented and compared with those obtained from Kaplan-Meier estimators and Cox regression. 

In Sections~\ref{sec:ULSAM} and \ref{sec:Amoris} our analysis is applied to real survival data related to prostate and breast cancer, from the ULSAM longitudinal cohort \cite{ULSAM,ULSAMpaper} and AMORIS study \cite{Holme_2010,Holme_2008}, respectively. The application to the ULSAM data leads to appealing and transparent new explanations for previously counter-intuitive inferences. Age-related survival differences between women diagnosed with breast cancer were found on application to data from the AMORIS cohort. In Section~\ref{sec:Discussion} we summarise our findings.

\section{Heterogeneity-induced informative censoring}
\label{sec:HICR}

In this section we introduce our survival analysis based on the assumption that risk event time correlations, if present, are caused by residual cohort heterogeneity. This assumption is much weaker than that of assuming risk independence, but it does allow us to overcome informative censoring and leads to an intuitive and transparent parametrisation of hazard rates, and makes quantities \emph{decontaminated} of its effects accessible.

The relationships between the event probabilities in a cohort and those of its members are critical to any analysis that seeks to address the issue of heterogeneity and are formalised in Section~\ref{ssec:CohortIndividualConnection}. In Section~\ref{ssec:RiskComplexity} the extent to which different risk (in)dependence assumptions limit survival analysis is explored; existing survival analysis methods and our heterogeneity-induced competing risk approach are compared in terms of the ``risk complexity'' of cohorts to which they can be applied. Decontaminated cause-specific survival functions and hazard rates are presented in Section~\ref{ssec:EffectsOfInformativeCensoring}, the difference between the decontaminated quantities and their respective crude counterparts providing a means of quantifying the severity of informative censoring in a cohort. The development of our latent class approach to modelling heterogeneous cohorts is outlined in Section~\ref{ssec:ModelGeneral} (with further details and identities given in Appendix~\ref{app:TheoryDetails}) and details of our Bayesian approach to the determination of the optimal characterisation of a cohort are given in Section~\ref{ssec:BayesianInference}. In Section~\ref{ssec:Implementation} the various outputs of our analysis and their benefits are discussed.

\subsection{Connection between cohort level and individual level descriptions}
\label{ssec:CohortIndividualConnection}

The standard mathematical relations of survival analysis are derived directly from the joint event time distribution and hold irrespective of whether we have a large or small cohort, or even a single individual. Below we formalise a number of quantities of interest, both for the individual and for the cohort as a whole, and the relationships between them; the index $i$ is used to denote those personalised quantities specific to individual $i$, with no such index being present for cohort-specific quantities.

We imagine a cohort of $N$ individuals who are each subject to $R$ true risks, labelled by $r=1\ldots R$, and an end-of-trial censoring event denoted by $r=0$\footnote{For the structure of the theory there is no difference between censoring due to alternative risks and censoring due to trial termination.}. The joint event time distribution $\Prob(t_0,\ldots,t_R)$ describes, for the cohort as a whole, the probability of the event times $(t_0,\ldots,t_R)$, where $t_r\geq 0$ is the time at which risk $r$ triggers an event\footnote{This starting point is not fully general. It assumes that all risks will ultimately lead to failure. One can include events with a finite chance of not happening at {\em any} time, by adding for each risk $r$ a binary variable $\tau_r$ to indicate whether or not the calamity button is pressed at time $t_r$.}, with the personalised event time distribution being denoted by $\Prob_i(t_0,\ldots,t_R)$. The cohort level risk event time distribution is a direct average of the personalised event risk event time distributions, such that $\Prob(t_0,\ldots,t_R)=N^{-1}\sum_{i}\Prob_i(t_0,\ldots,t_R)$. It is an estimator of the joint event time probability for the greater population if the cohort is representative. Throughout this paper we shall assume that $N$ is sufficiently large to ensure that the differences between $\Prob(t_0,\ldots,t_R)$ and the population's event time distribution are immaterial and that inferences made about the cohort are also true for the population.

In the interest of readability it should be assumed that products and summations over the risks run over the end-of-trial censoring risk and the true risks, such that $r=0,\dots,R$, shall be written as $\prod_{r}$ and $\sum_{r}$ respectively; should a particular risk $r^\prime$ be excluded then the product shall be written as $\prod_{r\neq r^\prime}$. Similarly, summations over the cohort shall run over all members, labelled by $i=1,\dots,N$, are written as $\sum_{i}$.

The crude cause-specific hazard rate, that is the probability per unit time that given failures occur at time $t$ if until then none of the possible events has yet occurred, follows from the joint event time distributions. The personalised cause-specific hazard rate, $h^i_r(t)$, is given by,

\begin{eqnarray}
h^i_r(t)&=&\frac{1}{S_i(t)}\int_0^\infty\!\!\!\!\ldots\!\int_0^\infty\!\dt_0\ldots \dt_R~\Prob_i(t_0,\ldots,t_R)\delta(t-t_r)\prod_{r^\prime\neq r}\theta(t_{r^\prime}-t),
\label{eq:rates_fori}
\end{eqnarray}

\noindent where the delta-distribution $\delta(x)$ is defined by the identity $\int_{-\infty}^\infty\!\rmd x~\delta(x)f(x)=f(0)$, the step function is defined as $\theta(x>0)=1$ and $\theta(x<0)=0$, and the personalised survival function $S_i(t)$ is given by $S_i(t)=\rme^{-\sum_{r} \int_0^t\ds~h^i_r(s)}$. The personalised probability density, $P_i(t,r)$, to find the earliest event occurring at time $t$ and corresponding to risk $r$ is given by $P_i(t,r)= h^i_r(t)\rme^{-\sum_{r^\prime} \int_0^t\ds~h^i_{r^\prime}(s)}$.

At the cohort-level, analogous relations define the cause-specific hazard rate, $h_r(t)$, survival function, $S(t)$, and time-risk event probability, $P(t,r)$, for the cohort as a whole; these relations are of identical form to the relations above though obviously cohort-level quantities replace individual-specific quantities. Typically, however, it is the characterisation of the cohort \emph{given} the survival analysis data for the cohort that is of interest; to achieve this we require the covariate-conditioned cause-specific hazard rates and survival functions. The covariate-conditioned risk event time probability distribution $\Prob(t_0,\ldots,t_R|\bz)$ describes the risk event time statistics of the sub-cohort of those individuals $i$ that have covariate vector $\bz_i=\bz$. At the cohort level, the covariate-conditioned cause-specific hazard rate is given by,

\begin{eqnarray}
h_r(t|\bz)=\frac{1}{S(t|\bz)}\int_0^\infty\!\!\!\!\ldots\!\int_0^\infty\!\dt_0\ldots \dt_R~\Prob(t_0,\ldots,t_R|\bz)\delta(t-t_r)\prod_{r^\prime\neq r}\theta(t_{r^\prime}-t),
\label{eq:covariates_h}
\end{eqnarray}

\noindent where the covariate-conditioned survival function is given by $S(t|\bz)=\rme^{-\sum_{r^\prime=0}^R\int_0^t\ds~h_{r^\prime}(s|\bz)}$. The covariate-conditioned risk event time probability density is then given by $P(t,r|\bz)=h_r(t|\bz)\rme^{-\sum_{r^\prime=0}^R\int_0^t\ds~h_{r^\prime}(s|\bz)}$.

The connection between the cohort-level and personalised risk event time distributions is simple; accounting for covariate-conditioning, the cohort level risk event time distribution is the average of the personalised event time distributions of those individials with $\bz_i=\bz$, such that $\Prob(t_0,\ldots,t_R|\bz)=\sum_{i,~\bz_i=\bz}\Prob_i(t_0,\ldots,t_R)/\sum_{i,~\bz_i=\bz}1$.

The relationship between quantities at the cohort and individual level is simple for those which depend linearly on the risk event time distribution. The cohort-level survival function is simply the average over the cohort of the individual survival functions, $S(t|\bz)=\sum_{i,~\bz_i=\bz}S_i(t)/\sum_{i,~\bz_i=\bz}1$ and the risk event time probability density is $P(t,r|\bz)=\sum_{i,~\bz_i=\bz}P_i(t,r)/\sum_{i,~\bz_i=\bz}1$. In contrast, quantities which depend on the risk event time distribution in a more complicated way, such as the crude cause-specific hazard rates, and cohort-level quantities are {\em not} direct averages over their individual level counterparts. The cohort level cause-specific hazard rates, for instance, are given by (see appendix \ref{app:link_for_rates} for details),

\begin{eqnarray}
h_r(t|\bz)&=&\frac{\sum_{i,\bz_i=\bz} h^i_r(t)\rme^{-\sum_{r^\prime=0}^R\int_0^t\ds~h^i_{r^\prime}(s)}}
{\sum_{i,\bz_i=\bz} \rme^{-\sum_{r^\prime=0}^R\int_0^t\ds~h^i_{r^\prime}(s)}}.
\label{eq:link2}
\end{eqnarray}

The cause-specific cumulative incidence function, $F_r(t)$, describes the probability that event $r$ has been {\em observed} at any time prior to time $t$,  and is given by,

\begin{eqnarray}
F_r(t)&=& \int_0^t\!\dt^\prime~S(t^\prime)h_r(t^\prime).
\end{eqnarray}

\noindent Although $F_r(t)$ refers to risk $r$ specifically, it can be heavily influenced by other risks. If it is small, this may be because event $r$ is intrinsically unlikely, or because it tends to be preceded by events $r^\prime\neq r$. One cannot tell.

\subsection{Risk complexity in heterogeneous cohorts}
\label{ssec:RiskComplexity}

In this section we explore the the consequences of event time risk correlations and cohort-level heterogeneity. One always allows cohorts to be heterogeneous in terms of covariates; we refer here to heterogeneity in the relation between covariates and risks.

A homogeneous cohort is one in which the relationship between an individual's covariates and risk is uniform throughout the cohort. In a homogeneous cohort the personalised event time distribution $\Prob_i(t_0,\ldots,t_R)$ can depend on $i$ only via $\bz_i$ as such there exists a function $\mathcal{Q}(t_0,\ldots,t_R,\bz)$ such that $\Prob_i(t_0,\ldots,t_R)=\mathcal{Q}(t_0,\ldots,t_R,\bz_i)$ is true for all individuals in the cohort. In such a cohort there can be no informative censoring and the crude and true cause-specific hazard rates and survival functions are identical.

In heterogeneous cohorts individuals have further relevant features which are not captured by their covariates. A consequence of these further features is that for such cohorts there no longer exists a function which, for all individuals in the cohort, maps between an individual's covariates and their personalised event time distribution. These additional features impact upon their risks. Here one will observe a gradual `filtering': high-risk individuals will drop out early, causing time dependencies at cohort level that have no counterpart at individual level. For instance, even if all individuals have stationary hazard rates, one would according to (\ref{eq:link2}) still find time dependent crude cohort level hazard rates. Here, having uncorrelated individual level risks no longer implies having uncorrelated  covariate-conditioned cohort level risks. 
It is quite possible to have ${\Prob}_i(t_0,\ldots,t_R)= \prod_{r}{\Prob}_{i}(t_r)$, but still ${\Prob}(t_0,\ldots,t_R|\bz)\neq \prod_{r}\Prob(t_r|\bz)$.

\begin{table}
\caption{Risk complexity to capture cohort heterogeneity, risk event-time correlations, and competing risks\label{tab:RiskComplexity}}
\vspace{3mm}
\centering
\begin{tabular}{llll}
\hline
\multicolumn{2}{l}{Risk complexity} & Individual & Cohort\\
\hline
\parbox[t]{0.3cm}{$~$\\I} &
\parbox[t]{5.75cm}{Homogenous cohort\\No competing risks at individual level\\No competing risks at cohort level} &
\parbox[t]{4.5cm}{$~$\\$\Prob_i(t_0,\ldots,t_R)=\prod_{r}\Prob(t_r|\bz_i)$} &
\parbox[t]{4.5cm}{$~$\\${\Prob}(t_0,\ldots,t_R|\bz)=\prod_{r}\Prob(t_r|\bz)$}
\\
\hline
\parbox[t]{0.3cm}{$~$\\II} &
\parbox[t]{5.75cm}{Heterogeneous cohort\\No competing risks at individual level\\No competing risks at cohort level} &
\parbox[t]{4.5cm}{$~$\\$\Prob_i(t_0,\ldots,t_R)=\prod_{r}\Prob_{i}(t_r)$} &
\parbox[t]{4.5cm}{$~$\\${\Prob}(t_0,\ldots,t_R|\bz)=\prod_{r}\Prob(t_r|\bz)$}
\\
\hline
\parbox[t]{0.3cm}{$~$\\III} &
\parbox[t]{6.0cm}{Heterogeneity-induced competing risks\\No competing risks at individual level\\Cohort level competing risks} &
\parbox[t]{4.5cm}{$~$\\$\Prob_i(t_0,\ldots,t_R)=\prod_{r}\Prob_{i}(t_r)$} &
\parbox[t]{4.5cm}{$~$\\${\Prob}(t_0,\ldots,t_R|\bz)\neq \prod_{r}\Prob(t_r|\bz)$}
\\
\hline
\parbox[t]{0.3cm}{$~$\\IV} &
\parbox[t]{5.75cm}{Heterogeneous cohort\\Individual level competing risks\\Cohort level competing risks} &
\parbox[t]{4.5cm}{$~$\\$\Prob_i(t_0,\ldots,t_R)\neq \prod_{r}\Prob_{i}(t_r)$} &
\parbox[t]{4.5cm}{$~$\\${\Prob}(t_0,\ldots,t_R|\bz)\neq \prod_{r}\Prob(t_r|\bz)$}\\
\hline
\end{tabular}
\end{table}

Risk event time correlations in a cohort can be generated at different levels. There is a natural hierarchy of cohorts in terms of risk complexity, as summarised in Table~\ref{tab:RiskComplexity}, with implications for the applicability of the different survival analysis methods.

Assuming statistically independent risk event times at cohort level underlies both Kaplan-Meier estimators and Cox regression, and is seen only in those cohorts with risk complexity levels I and II. At level II there is still no competing risk problem, but heterogeneity may demand parametrisations of crude cohort level primary hazard rates that could be more complex than those of Cox, which is the rationale behind frailty and random effects models, and the latent class models of \cite{Muhten}. All these approaches still only model the primary risk, and therefore cannot handle cohorts beyond level II.

In this paper we shall focus on developing survival analysis tools to investigate cohorts with risk complexity level III, in which the event times of all risks are assumed to be statistically independent for each individual but residual heterogeneity leads to risk correlations at cohort level. In such cohorts the correlations between cohort level event times  have their origin strictly in {\em correlations between disease susceptibilities and covariate associations of individuals}, for example, someone with a high hazard rate for a disease $A$ may also be likely to have a high hazard rate for event $B$, for reasons not explained by the covariates. At this level of risk complexity competing risks phenomena shall be observed as the risk correlations at cohort level will cause informative censoring.

Cohorts of risk complexity level IV are the most complex and difficult to model, with the event times of different risks assumed to be correlated at both individual and cohort level, and shall not be explored in this paper.

\subsection{Separating direct from indirect associations and quantifying informative censoring}
\label{ssec:EffectsOfInformativeCensoring}

The assumption of heterogeneity-induced competing risks, being that the event times of all risks are statistically independent for each individual in the cohort but not for the cohort as a whole (risk complexity level III in our classification above), allows us to investigate \emph{analytically} the effects of informative censoring. This additional assumption addresses the issue of identifiabilty \cite{Tsiatis} and, with the risk event time marginal distributions now accessible, it becomes possible to develop expressions not only for the crude cause-specific hazard rates and survival function but also for their decontaminated counterparts.

The personalised cause-specific event time probability given the assumption of risk independence at the individual level is given by $\Prob_{i}(t_r)=h_r^i(t){\rm e}^{-\int_0^t\ds~h_r^i(s)}$. The covariate-conditioned cohort-level risk event time marginals are therefore given by $\Prob(t_r|\bz)=\sum_{i,\bz_i=\bz}h_r^i(t){\rm e}^{-\int_0^t\ds~h_r^i(s)}/\sum_{i,\bz_i=\bz}1$, and can be used to develop expressions for the decontaminated survival functions and hazard rates\footnote{The decontaminated survival functions and hazard rates follow from $\tilde{S}_r(t|\bz)=\int_t^\infty\!\!\dt_r~\Prob(t_r|\bz)$ and $\tilde{h}_r(t|\bz)=-{\rm\frac{d}{\dt}}\log \tilde{S}_r(t|\bz)$.}.

The crude cause-specific hazard rates which would have been found if all risks had been independent, $h_r(t|\bz)$, and their decontaminated counterparts, $\tilde{h}_r(t|\bz)$, are given by,

\begin{eqnarray}
h_r(t|\bz)=\frac{\sum_{i,\bz_i=\bz} h^i_r(t)\rme^{-\sum_{r^\prime=0}^R\int_0^t\ds~h^i_{r^\prime}(s)}}
{\sum_{i,\bz_i=\bz} \rme^{-\sum_{r^\prime=0}^R\int_0^t\ds~h^i_{r^\prime}(s)}}
\label{eq:falseh}
,\qquad
\tilde{h}_r(t|\bz)=
\frac{\sum_{i,\bz_i=\bz}h_r^i(t)\rme^{-\int_0^{t}\ds~h_r^i(s)}}{\sum_{i,\bz_i=\bz}\rme^{-\int_0^{t}\ds~h_r^i(s)}}.
\label{eq:truehtilde}
\end{eqnarray}

\noindent The crude and decontaminated cause-specific hazard rates will generally have different values. In the decontaminated cause-specific hazard rate, $\tilde{h}_r(t|\bz)$, the probability that individual $i$ survives until time $t$ is given by $\exp[-\int_0^{t}\ds~h_r^i(s)]$. The probability of survival until time $t$ for individual $i$ in the crude cause-specific hazard rate, $h_r(t|\bz)$, depends on \emph{all} risks.

The crude cause-specific survival function, $S_r(t|\bz)$, and its decontaminated counterpart, $\tilde{S}_r(t|\bz)$, are given by,

\begin{eqnarray}
S_r(t|\bz)= \rme^{-\int_0^t\ds~h_r(s|\bz)}
\label{eq:falseS}
,\qquad
\tilde{S}_r(t|\bz)=
\frac{\sum_{i,\bz_i=\bz}\rme^{-\int_0^{t}\ds~h_r^i(s)}}
{\sum_{i,\bz_i=\bz}1}.
\label{eq:trueStilde}
\end{eqnarray}

The decontaminated cause-specific survival function, $\tilde{S}_r(t|\bz)$, depends only on the risk $r$, whereas its crude counterpart depends on \emph{all} risks through the crude hazard rate $h_r(t|\bz)$.

Hence the assumption that competing risks (if present) are induced by heterogeneity leads to relatively simple formulae for the decontaminated cause-specific quantities of interest
and for the likelihood of observing individual survival data. What remains is to identify the {\em minimal} level of description required for evaluating these formulae, and to determine how the required information can  be estimated from survival data.

\subsection{Modelling the heterogeneous cohort}
\label{ssec:ModelGeneral}

In this section we shall develop our latent class model. At the heart of our model are the personalised cause-specific hazard rates, $h_r^i(t)$, being of the Cox form and obeying the assumption of proportional hazards.

In constructing our latent class framework, we assume a heterogeneous cohort to be comprised of $L$ sub-cohorts, or latent classes, labelled by $\ell=1,\dots,L$. Each individual from the cohort must belong to one of the latent classes; as such the classes are discrete, no class can contain a member from any other class, and together the classes contain all of the individuals in the cohort. Each class obeys the proportional hazards assumption although the cohort collectively need not.

The personalised cause-specific hazard rates shall depend on cause-specific and possibly class-specific frailty, association, and base hazard rate parameters. The frailty parameters, $\beta_r^{\ell 0}$, capture the impact of unobserved variables within class $\ell$ for risk $r$, i.e. effects that cannot be attributed to the included covariates directly, the association parameters, $\beta_r^{\ell \mu}$, quantify how strongly each of the $\mu=1,\dots,P$ covariates influence the personalised hazard rate, and the base hazard rate, $\lambda_{r}^{\ell}(t)$, describes the cause-specific personalised hazard rate for an individual from class $\ell$ having exactly \emph{average} covariate values. Inevitably, modelling {\em all} risks and their correlations leads to models having more parameters than those which model \emph{only} the primary risk. In an effort to avoid overfitting, three variants of the personalised cause-specific hazard rates are introduced, with differing degrees of heterogeneity, as summarised in Table~\ref{tab:ModelVariants}.

\begin{table}
\caption{{Personalised cause-specific hazard rates, for each of the $R$ active risks, to capture latent cohort heterogeneity}\label{tab:ModelVariants}}
\vspace{3mm}
\centering
\begin{tabular}{lll}
\hline
\parbox[t]{1.1cm}{$~$\\$M=1$} &
\parbox[t]{5.3cm}{Heterogeneous frailties\\Homogeneous associations\\Homogeneous base hazard rates} &
\parbox[t]{4.5cm}{$~$\\$h_r^i(t)=\lambda_{r}(t) {\rm e}^{\beta_{r}^{\ell 0} + \sum_{\mu}\beta_{r}^{\mu}z_i^{\mu} }$}\\\hline
\parbox[t]{1.1cm}{$~$\\$M=2$} &
\parbox[t]{5.3cm}{Heterogeneous frailties\\Heterogeneous associations\\Homogeneous base hazard rates} &
\parbox[t]{4.5cm}{$~$\\$h_r^i(t)=\lambda_{r}(t) {\rm e}^{\beta_{r}^{\ell 0} + \sum_{\mu}\beta_{r}^{\ell\mu}z_i^{\mu} }$}
\\
\hline
\parbox[t]{1.1cm}{$~$\\$M=3$} &
\parbox[t]{5.3cm}{Heterogeneous frailties\\Heterogeneous associations\\Heterogeneous base hazard rates} &
\parbox[t]{4.5cm}{$~$\\$h_r^i(t)=\lambda_{r}^{\ell}(t) {\rm e}^{\beta_{r}^{\ell 0} + \sum_{\mu}\beta_{r}^{\ell\mu}z_i^{\mu} }$} 
\\
\hline
\end{tabular}
\end{table}

When applying our latent class framework to describe cohort-level quantities of interest, such as the cause-specific survival and cumulative incidence functions, those quantities involving the individualised hazard rates will be expressed in terms of the variables of the latent class model; summations over the individuals in the cohort become summations over the latent classes, the relative influence of each class depending on the fraction of individuals from the cohort belonging to that class. Representing the collection of those individuals belonging to class $\ell$ by the variable $I_\ell$, it follows that a summation over all individuals in the cohort, $\sum_{i,\bz_i=\bz}$, becomes a summation over the latent classes \emph{and} their membership, of the form $\sum_{\ell}\sum_{i \in I_\ell, \bz_i=\bz}$. Those quantities expressed through summations over the individualised cause-specific hazard rates, $h_r^i(t)$, can then be written as summations over covariate-conditioned class- and cause-specific hazard rates, $h_r^\ell(t|\bz)$, as follows,

\begin{eqnarray}
\frac{\sum_{i,\bz_i=\bz} f(h_r^i(t))}{\sum_{i,\bz_i=\bz} 1} =
\frac{\sum_{\ell}\sum_{i \in I_\ell, \bz_i=\bz} f(h_r^\ell(t|\bz))}{\sum_{\ell}\sum_{i \in I_\ell, \bz_i=\bz} 1} =
\frac{\sum_{\ell} (n_\ell(\bz)/p(\bz)N) f(h_r^\ell(t|\bz))}{\sum_{\ell} (n_\ell(\bz)/p(\bz)N)} =
\sum_{\ell} w_\ell(\bz) f(h_r^\ell(t|\bz)),
\end{eqnarray}

\noindent where the class membership fraction, $w_\ell(\bz)$, given by the quantity $n_\ell(\bz)/p(\bz)N$, follows from the fraction of individuals from the cohort who belong to class $\ell$ and have covariates $\bz$ (where $p(\bz)$ represents the population probability of the covariate vector $\bz_i$ being equal to $\bz$). The class membership fractions are subject to the constraint $\sum_\ell w_\ell(\bz)=1$. In this paper, to avoid further complexity the class membership fractions are chosen to be independent of the covariates, this amounts to the assumption that all sub-cohorts have identically distributed covariates.

The various crude and decontaminated quantities which describe a cohort are given below in terms of our latent class parametrisation for the fully heterogeneous variant of the individualised cause-specific hazard rate ($M=3$); corresponding expressions are easily obtained for the simpler model variants on substitution of class-independent association(s) ($M=1$) and class-independent base hazard rate(s) ($M=2$) as appropriate. To aid readability the following compact notation is used: the effect of the frailty and associations for each true risk $r$ and class $\ell$ are summarised by the product $\bbeta_r^\ell\cdot\bz=\beta_r^{\ell 0}+\sum_{\mu}\beta_r^{\ell\mu} z^\mu$, the time integrals can be factorised and denoted by $\Lambda_r^\ell(t)=\int_0^{t}\ds~\lambda_r^\ell(s)$ as the base hazard rates contain \emph{all} time dependencies. The end-of-trial risk is assumed not to depend on the covariates, accordingly the class-independent associations for the censoring risk are defined to be zero $\bbeta_0=0$. Summations over the covariates run over all $P$ covariates, labelled by $\mu=1,\dots,P$, and shall be written as $\sum_{\mu}$.

In their parametrised form, the crude and decontaminated hazard rate expressions for the true risks $r=1,\dots,R$ are given by,

\begin{eqnarray}
h_r(t|\bz)=
\frac{\sum_{\ell} w_\ell ~{\lambda}_r^{\ell}(t)\rme^{{\bbeta}^\ell_r\cdot \bz-\sum_{r^\prime=1}^R \exp({\bbeta}^\ell_{r^\prime}\cdot \bz){\Lambda}_{r^\prime}^{\ell}(t)}}
{\sum_{\ell} w_\ell ~\rme^{-\sum_{r^\prime=1}^R \exp({\bbeta}^\ell_{r^\prime}\cdot \bz){\Lambda}_{r^\prime}^{\ell}(t)}},
\label{eq:cruderateLC}
\qquad
\tilde{h}_r(t|\bz)=
\frac{ \sum_{\ell} w_\ell  ~{\lambda}_r^{\ell}(t)\rme^{{\bbeta}^\ell_r\cdot \bz- \exp(\hat{\bbeta}^\ell_{r}\cdot \bz){\Lambda}_{r}^{\ell}(t)}}
{\sum_{\ell} w_\ell  ~\rme^{- \exp({\bbeta}^\ell_{r}\cdot \bz){\Lambda}_{r}^{\ell}(t)}}.
\end{eqnarray}

\noindent The corresponding crude and decontaminated survival functions are given by,

\begin{eqnarray}
S_r(t|\bz)=
\exp\left(-\int_0^t\!\rmd s~ h_r(s|\bz)\right),
\qquad
\tilde{S}_r(t|\bz)=
\sum_{\ell} w_\ell  ~\rme^{- \exp({\bbeta}^\ell_{r}\cdot \bz){\Lambda}_{r}^{\ell}(t)}.
\label{eq:StildeLC}
\end{eqnarray}

\noindent Once more, it is clear that as the crude cause-specific hazard rate is influenced by \emph{all} risks then the crude cause-specific survival function shall be also. In Appendix~\ref{app:CrudeDeconNoRiskCorrelations} the crude and decontaminated survival are shown to be identical in the case where there is one risk only, i.e. in the absence of any potential informative censoring.

The cause-specific cumulative incidence functions using our latent class model are given by,

\begin{eqnarray}
F_r(t|\bz)= 
\int_0^t\!\!\rmd t^\prime~\rme^{-{\Lambda}_0(t^\prime)}~
\sum_{\ell} w_\ell ~{\lambda}_r^{\ell}(t^\prime)
\rme^{{\bbeta}^\ell_r\cdot \bz-\sum_{r^\prime=1}^R \exp({\bbeta}^\ell_{r^\prime}\cdot \bz){\Lambda}_{r^\prime}^{\ell}(t^\prime)}.
\label{eq:CumulativeIncidence}
\end{eqnarray}

\noindent It is noteworthy that our approach leads to an intuitive and easily interpreted formulation of the cause-specific cumulative incidence in which the role of all model parameters is completely transparent.

Our approach also offers \emph{retrospective} determination of class membership probability for an individual using their covariates, $\bz$, and survival information, $(t,r)$. Following Bayesian arguments (as detailed in Appendix~\ref{app:ClassAssignment}), the probability that individual belongs to class $\ell$, given their covariates and survival information, is given by,

\begin{eqnarray}
P(\ell|t,r,\bz)=
\frac{ w_{\ell}~\lambda_{r}^{\ell}(t) \rme^{\bbeta_{r}^{\ell}\cdot\bz-\sum_{r^\prime=1}^{R} \exp\left(\bbeta_{r^\prime}^{\ell}\cdot\bz\right)\Lambda_{r^\prime}^{\ell}(t)}}
{\sum_{\ell^\prime=1}^L w_{\ell^\prime} ~\lambda_{r}^{\ell^\prime}(t) \rme^{\bbeta_r^{\ell^\prime}\cdot\bz-\sum_{r^\prime=1}^R \exp\left(\bbeta_{r^\prime}^{\ell^\prime}\cdot\bz\right)\Lambda_{r^\prime}^{\ell^\prime}(t)}}.
\label{eq:find_class}
\end{eqnarray}

\noindent The search for informative new covariates could be aided by retrospective class assignment, increasing our ability to predict personalised risk in heterogeneous cohorts. Such new covariates are expected to be features that patients in the same class have in common.

Determination of the optimal characterisation of a cohort, that is the optimal values of the latent class fractions and the frailties, associations, and base hazard rates for each class and each risk, given the available data, is described in the next section.

\subsection{Characterisation of cohort heterogeneity}
\label{ssec:BayesianInference}

The Bayesian formalism has been used to determine the optimal characterisation of a cohort. The most appropriate latent class model variant, the most suitable number of latent classes, and the form of the base hazard rates are all unknown at the outset of an analysis, so their optimal values and the corresponding optimal class membership weightings, frailties, associations, and base hazard rates all need to be determined.

The term `model' shall be used to describe the combination of a particular number of latent classes, $L$, a particular form of the base hazard rates realised by a spline construction having $K$ anchored time points, and a particular parametrisation of the personalised hazard rates, $M$, and shall be denoted by $\mathcal{H}_{KLM}$. The parameter vector, $\btheta_{\mathcal{H}_{KLM}}$, shall denote the model parameters (i.e. the class membership fractions, the frailty, association, and base hazard rate parameters) of the model $\mathcal{H}_{KLM}$. The expressions developed below apply for the fully heterogeneous personalised cause-specific hazard rate ($M=3$); equivalent expressions for the simpler model variants ($M=1$, $M=2$) are easily obtained on substituting class-independent base hazard rate(s) and association(s) as appropriate.

The probability density to find the earliest event occurring at time $t$ corresponding to risk $r$ provides the link between our latent class model parameters and the observed survival data. The probability density for an individual with covariate vector $\bz$ to report $(t,r)$ is given by,

\begin{eqnarray}
P(t,r|\bz)= 
\rme^{-{\Lambda}_0(t)}\sum_{\ell} w_\ell ~{\lambda}_r^{\ell}(t) \rme^{{\bbeta}^\ell_r\cdot \bz-\sum_{r^\prime=1}^R \exp({\bbeta}^\ell_{r^\prime}\cdot \bz){\Lambda}_{r^\prime}^{\ell}(t)},
\label{eq:RiskTimeProb}
\end{eqnarray}

\noindent where the parameters are those familiar from the fully-heterogeneous personalised hazard rate ($M=3$) defined in Table~\ref{tab:ModelVariants}. The data likelihood, $P(D|\btheta_{\mathcal{H}_{KLM}})= \prod_{i} P(t_i,r_i|\bz_i,\btheta_{\mathcal{H}_{KLM}})$, gives the joint probability of the event times $t_i$ and risk $r_i$ for every individual in the cohort, conditioned on their covariates, $\bz_i$, and the model parameters, $\btheta_{\mathcal{H}_{KLM}}$, for the model $\mathcal{H}_{KLM}$. The task of identifying the optimal characterisation of a cohort is clear; it is necessary to find the most probable parameter values, $\btheta_{\mathcal{H}_{KLM}}^\star$, of the most probable model, $\mathcal{H}_{KLM}^{\star}$.

The Bayesian formalism is first applied to determine the optimal parameter values, $\btheta_{\mathcal{H}_{KLM}}^\star$, for each model $\mathcal{H}_{KLM}$. The posterior distribution gives the probability of the model parameters, $\btheta_{\mathcal{H}_{KLM}}$, conditioned on the cohort survival data, $D$, and is given by $P(\btheta_{\mathcal{H}_{KLM}}|D) = Z_{\mathcal{H}_{KLM}}^{-1} \exp(\LL(\btheta_{\mathcal{H}_{KLM}},D))$, where the log-likelihood is defined as $\LL(\btheta_{\mathcal{H}_{KLM}},D) = \log(P(D|\btheta_{\mathcal{H}_{KLM}})P(\btheta_{\mathcal{H}_{KLM}}))$, the normalisation constant is denoted by $Z_{\mathcal{H}_{KLM}}$, and the prior distribution over the model parameters is given by $P(\btheta_{\mathcal{H}_{KLM}})$ (Appendix~\ref{app:ModelPriors}). The contribution to the log-likelihood from the data-likelihood, for the fully heterogeneous model variant ($M=3$), is given by,

\begin{eqnarray}
\log P(D|\btheta_{\mathcal{H}_{KLM}}) =
-\sum_{i} \Lambda_0(t_i)
+
\sum_{i} \log \left[\sum_{\ell} w_\ell ~{\lambda}_{r_i}^{\ell}(t_i)
\rme^{{\bbeta}_{r_i}^{\ell}\cdot\bz_i-\sum_{r=1}^R\! {\Lambda}_{r}^{\ell}(t_i)\exp({\bbeta}_{r}^{\ell}\cdot\bz_i)}
\right].
\end{eqnarray}

\noindent In this work, maximum a posteriori parameter estimation (MAP) has been used; the optimal parameter values being those which maximise the posterior probability.

Bayesian model selection is used to determine the most probable model, $\mathcal{H}_{KLM}^\star$, given the cohort survival data, $D$, from an \emph{ensemble} of models\footnote{The ensemble of models is assumed to contain the actual model underlying the \emph{true} cohort sub-structure.}, where all models $\mathcal{H}_{KLM}$ are given identical prior probability $P(\mathcal{H}_{KLM})$. In pursuing this approach, the optimal model is that which is supported by the greatest ``evidence" $P(\mathcal{H}_{KLM}|D)$, which is here proportional to the posterior normalisation constant $Z_{\mathcal{H}_{KLM}}=\int\!\rmd\btheta_{\mathcal{H}_{KLM}}~P(D|\btheta_{\mathcal{H}_{KLM}},\mathcal{H}_{KLM})P(\btheta_{\mathcal{H}_{KLM}}|\mathcal{H}_{KLM})$ for that model. In this work a Gaussian approximation to the posterior has been used such that $Z_{\mathcal{H}_{KLM}}$ may be determined without the requirement for computationally expensive numerical integration (Appendix~\ref{app:ModelEvidenceGaussianApprox}), but can be determined from the optimal model parameter values, $\btheta_{\mathcal{H}_{KLM}}^\star$, and shall be written as $Z_{\mathcal{H}_{KLM}} = Z(\btheta_{\mathcal{H}_{KLM}}^\star)$.

In practice, to find the optimal characterisation of a cohort, $\btheta_{\mathcal{H}_{KLM}^\star}^\star$, requires that the optimal parameter values for each of the models being compared are \emph{first} determined and \emph{then} the model supported by the greatest evidence is identified. The optimal parameter values, $\btheta_{\mathcal{H}_{KLM}}^\star$, are determined for every model before the optimal model, $\mathcal{H}_{KLM}^\star$, from the ensemble is determined, according to the following algorithm,

\begin{eqnarray}
\mathcal{H}_{KLM}^\star = {\argmax}_{\mathcal{H}_{KLM}}\left[ Z(\btheta_{\mathcal{H}_{KLM}}^\star)\right], \qquad \btheta_{\mathcal{H}_{KLM}}^\star={\argmax}_{\btheta_{\mathcal{H}_{KLM}}} \left[P(D|\btheta_{\mathcal{H}_{KLM}})P(\btheta_{\mathcal{H}_{KLM}})\right].
\label{eq:OptimalModelAndParams}
\end{eqnarray}

\subsection{Practical tools for survival analysis}
\label{ssec:Implementation}

Implementation of the analysis protocol described above (Section~\ref{ssec:BayesianInference}) has been realised in our software package, \emph{ALPACA} (\textbf{A}dvanced \textbf{L}atent Class \textbf{P}rediction \textbf{A}nd \textbf{C}ompeting Risk \textbf{A}nalysis), using the C programming language. The optimal frailty, association, and base hazard rate parameters for each model are located by MAP estimation using a stochastic refinement of the downhill simplex method \cite{NR}. Numerical estimation of the curvature of the posterior distribution around the location of maximum probability enables both the error bars for the parameter estimates and the model ``evidence" to be determined. As the search for the optimal parameter values is achieved using a stochastic optimization algorithm, and as the task of searching the parameter space becomes more difficult as the number of latent classes is increased and the greater the complexity of the personalised hazard rate model and base hazard rate, this procedure is typically performed multiple times and the overall optimal characterisation is selected.

A cohort's survival data is pre-processed before application of our analysis algorithms. This pre-processing involves the linear rescaling of the \emph{raw} covariate values such that the distribution of their \emph{transformed} counterparts correspond to a zero average and unit variance distribution, equivalent to the definition of $Z$-scores. Missing data values are imputed according to the average of those values which are available. The parameter estimates obtained using our analysis can be translated to the more familiar language of hazard ratios (HR), 95\% confidence intervals (CI), and $p$-values, as a consequence of such data pre-processing. An additional benefit of this pre-processing is that the magnitudes of the effects for different associations can be directly compared.

The hazard ratio, ${\rm{HR}}_\mu$, associated with covariate $\mu$, follows from generalising the concept to a balanced cohort\footnote{It is not possible to capture risk in a single number for non-binary covariates.}, and can, as a consequence of our normalisation of covariates, be computed from the estimated association parameter $\beta_\mu$ according to the relation ${\rm{HR}}_\mu = \rme^{2\beta_\mu}$. The lower and upper bounds, ${\rm{HR}}_\mu^{-}$ and ${\rm{HR}}_\mu^{+}$ respectively, of the 95\% confidence interval, ${\rm{CI}}_\mu = [{\rm{HR}}_\mu^{-},{\rm{HR}}_\mu^{+}]$, for the hazard ratio associated with covariate $\mu$ are related to the estimated association parameter $\beta_\mu$ and its uncertainty $\sigma_\mu$ by the expressions ${\rm{HR}}_\mu^{-}=\exp(2(\beta_\mu-1.96\sigma_\mu))$ and ${\rm{HR}}_\mu^{+}=\exp(2(\beta_\mu+1.96\sigma_\mu))$ respectively. The $p$-value is approximated from our parameter estimates according to the relation $p_\mu=1-\erf(|\beta_\mu|/\sqrt{2}\sigma_\mu)$ where $|\beta_\mu|$ denotes the magnitude of the estimated association parameter and the error integral is given by $\erf(t)=(2/\sqrt{\pi})\int_0^t~\rmd  x ~\rme^{-x^2}$.

Survival curves, cumulative incidence curves, and retrospective class allocation for each individual are all possible once the optimal frailty, association, and base hazard rate parameters have been determined. The presence of informative censoring in a cohort can be deduced from observing differences between the crude and decontaminated survival curves; the magnitude of any such differences being indicative of the extent to which competing risks mask the true cause-specific survival. False protectivity effects should be suspected if the crude survival function exceeds the decontaminated survival function; the opposite being suggestive of the influence of false exposure in the cohort data.

In a heterogeneous cohort, the class-specific decontaminated survival may offer valuable insight into the expected progression of individuals belonging to each of the latent classes. Expressing the decontaminated cause-specific survival (\ref{eq:StildeLC}), $\tilde{S}_r(t|\bz)$, as the weighted sum of cause- and class-specific survival, $\tilde{S}_r^{\ell}(t|\bz)$, as given by,

\begin{eqnarray}
\tilde{S}_r^{\ell}(t|\bz)=
\rme^{- \exp({\bbeta}^\ell_{r}\cdot \bz){\Lambda}_{r}^{\ell}(t)},
\label{eq:StildeClassSpecific}
\end{eqnarray}

\noindent allows the class-specific decontaminated survival for each of the latent classes to be compared. Similarly, the cause-specific cumulative incidence function (\ref{eq:CumulativeIncidence}) can be expressed as a weighted sum of its class-specific components, $F_r^\ell(t|\bz)$, given by,

\begin{eqnarray}
F_r^\ell(t|\bz)= 
\int_0^t\!\!\rmd t^\prime~\rme^{-\hat{\Lambda}_0(t^\prime)}~ \hat{\lambda}_r^{\ell}(t^\prime)
\rme^{\hat{\bbeta}^\ell_r\cdot \bz-\sum_{r^\prime=1}^R \exp(\hat{\bbeta}^\ell_{r^\prime}\cdot \bz)\hat{\Lambda}_{r^\prime}^{\ell}(t^\prime)},
\label{eq:CumulativeIncidenceClassSpecific}
\end{eqnarray}

\noindent such that $F_r(t|\bz)=\sum_{\ell} w_\ell F_r^\ell(t|\bz)$. The class- and cause-specific cumulative incidence provides information as to the relative occurrence of events for each latent class and each cause at any time during the trial.

Retrospective class assignment, by identification of the most probable latent class to which an individual belongs using (\ref{eq:find_class}), can offer additional insight into differences between latent classes. Retrospectively allocated class-conditioned time-to-event distributions may offer clues as to the expected survival time for members of the different classes. The detection of differences between covariate distributions for those individuals retrospectively assigned to each latent class offers a potentially powerful means of identifying novel informative covariates.

\section{Application to synthetic survival data}
\label{sec:SynthData}

In this section the results of the application of our algorithms to synthetic data (see Appendix~\ref{app:SynthData}) simulating a variety of conditions are presented. The effectiveness of our algorithm to successfully characterise a heterogeneous cohort having three latent classes, two of which differ only in their base hazard rates is demonstrated in Section~\ref{ssec:SynthData_CohortSubstructure}. In Section~\ref{ssec:SynthData_InformativeCensoring} the ability of our analysis to accurately characterise a cohort in the presence of heterogeneity-induced informative censoring is shown, the differences between the crude and decontaminated survival curves indicating the influence of competing risks.

The model space for the analysis of each synthetic data set covered \emph{all} combinations of latent classes between one and four, base hazard rate complexity between one and eight, and for each of the three variants of the personalised hazard rate. The optimal parameter estimates were obtained for each model at least five times and the model with the greatest overall evidence identified to provide the optimal characterisation for each of the synthetic data sets.

As the class membership of all individuals is known when using simulated cohort data it is also possible to quantify the accuracy of Bayesian retrospective class allocation; the effectiveness of class identification using our algorithm (\ref{eq:find_class}) is quantified by calculating the fraction of correctly assigned individuals and the quality of allocation can be expressed for each class, $q_\ell$, or for the cohort as a whole, $q$. It is important to note that, even if all parameters were known exactly, due to the stochasticity of event times class allocation will never be perfectly accurate\footnote{An upper bound on the possible accuracy of retrospective allocation can be determined using the retrospective allocation and event time probability relations under the assumption of an infinitely large cohort from which the parameters have been extracted perfectly; for a cohort subject to one risk and consisting of two latent classes of equal size, with a constant class-independent base hazard rate and only one covariate, the best possible allocation quality in the case that $\beta_1^{11}=-\beta_1^{21}=2$ is about 83\%.}.

\subsection{Revealing heterogeneity and cohort sub-structure}
\label{ssec:SynthData_CohortSubstructure}

Our analysis is now applied to data modelling a heterogeneous cohort having three latent classes ($L=3$) but free of the effects of informative censoring. The characteristics of \emph{Cohort A} are given in Table~\ref{tab:SynthDataA}; the three classes are of equal size, each individual $i$ has three covariates $(z_i^1,z_i^2,z_i^3)$ and is subject to either one real risk or end-of-trial censoring at time $t=20$. The personalised cause-specific hazard rates of individuals in \emph{Cohort A} have class-dependent associations and base hazard rates (i.e. are of type $M=3$). The associations of individuals belonging to classes $\ell=1$ and $\ell=2$ are identical; the classes differ only in their base hazard rates, the base hazard rate of class $\ell=1$ is time-independent, the base hazard rate of class $\ell=2$ increases exponentially with time. The base hazard rate of the third class $\ell=3$ is time-independent, though it differs from that of the first class.

\begin{table}
\caption{{Modelling a heterogeneous cohort:} The parameters used to generate synthetic data modelling a cohort having three latent classes, one real risk, and with end-of-trial censoring at time $t=20$. The frailty parameter, $\beta_r^{\ell 0}$, for each latent class has been set to zero. \label{tab:SynthDataA}}
\vspace{3mm}
\centering
\begin{tabular}{lrrr}
\hline
\multicolumn{4}{l}{\emph{Cohort A: ($L=3,M=3$)}}\\
\multicolumn{4}{l}{Heterogeneity and class-dependent base hazard rates}\\
\hline
 & Class, $\ell=1$ & Class, $\ell=2$ & Class, $\ell=3$ \\
\hline
$w_\ell$ & 1/3 & 1/3 & 1/3 \\
\hline
$\lambda_1^\ell(t)$ & 3/10 & $\rme^{t/4}/100$ & 1/10\\
\hline
$\beta_1^{\ell 1}$ & 2 & 2 & -2\\
$\beta_1^{\ell 2}$ & 0 & 0 & 0 \\
$\beta_1^{\ell 3}$ & 0 & 0 & 0 \\
\hline
\end{tabular}
\end{table}

\begin{figure}[t]
\centering
\includegraphics{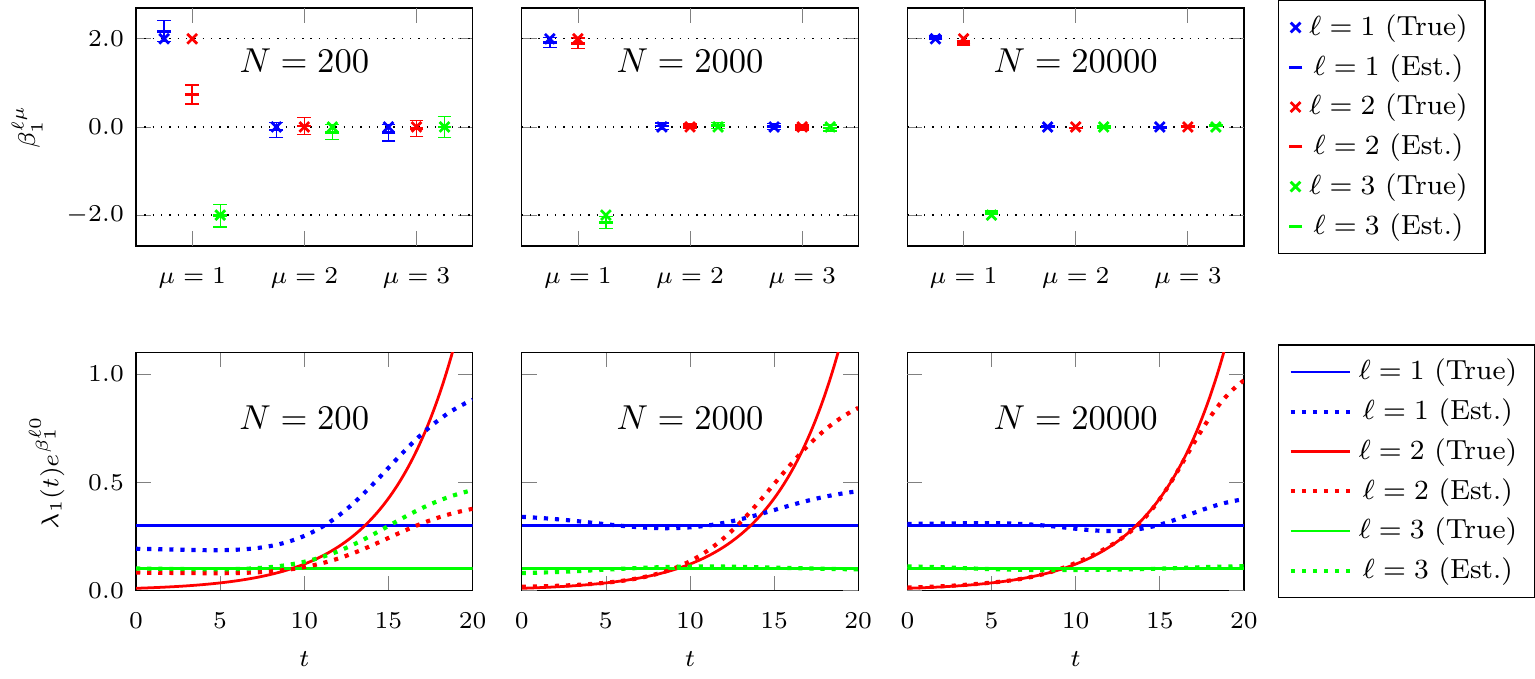}
\caption{\footnotesize{\emph{Bayesian identification and characterisation of a synthetic heterogeneous cohort:} The optimal association and base hazard rate estimates for analysis of \emph{Cohort A} data, as specified in Table~\ref{tab:SynthDataA}, are shown along with their true values for sample sizes $N=200$, $N=2000$, and $N=20000$. Our analysis correctly identifies three latent classes even for the meagre sample size of $N=200$; though it is clear on inspection of the estimates that their accuracy and precision improves with increasing sample size. The association parameters are accurately estimated for all covariates and all classes on analysis of \emph{Cohort A} data of size $N=2000$; the base hazard rates deviate only minimally from their true value at times greater than about $t=10$. The precision of the estimated associations is superior for the data set of size $N=20000$ and the estimated base hazard rates are true to their actual values even beyond time $t=15$.}
}
\label{fig:DataA_ParamEstimates}
\end{figure}

The characterisation of such a cohort is challenging even in the absence of informative censoring; to successfully characterise the cohort \emph{any} analysis must be able to identify the three latent classes and distinguish between two classes which vary from each other only in their base hazard rate. The effectiveness of our analysis to characterise \emph{Cohort A} is summarised in Figure~\ref{fig:DataA_ParamEstimates}, which shows the optimal estimated associations and base hazard rates for data set sizes $N=20000$, $N=2000$, and $N=200$. It is evident (Figure~\ref{fig:DataA_ParamEstimates}) that the association parameters of \emph{Cohort A} were accurately estimated for data set sizes $N=2000$ and $N=20000$, and that the estimated and true base hazard rates are sufficiently close that the optimal model can be considered as providing a good characterisation of the cohort structure. Although it is clear that a sample size of $N=200$ is insufficient for \emph{Cohort A} to be accurately characterised, it is noteworthy that despite such a meagre sample size our analysis correctly reports that the cohort is comprised of three latent classes and the estimated associations are accurate (within the estimated uncertainty) for covariates 2 and 3 for all three classes, and are accurate for classes 1 and 3 for covariate 1.

The optimal models reported for data set sizes $N=20000$ and $N=2000$ were $\mathcal{H}_{KLM}^\star=(L=3,M=3,K=4)$ and $\mathcal{H}_{KLM}^\star=(K=3,L=3,M=3)$ respectively. In total, 23 parameters (two weights, and for each of the three latent classes one frailty, three associations, and three modifiable spline points for the parametrised base hazard rate approximation) were estimated for the optimal model for the $N=20000$ data set. As the Bayesian model selection stage of our analysis seeks to balance the justification for greater model complexity against the available data, it is unsurprising that models containing fewer parameters were found to be optimal for the $N=2000$ and $N=200$ data sets, having 20 and 16 parameters respectively. Ultimately, there is not enough information available in the $N=200$ data set to justify the selection of a model having the required complexity to accurately describe \emph{Cohort A} and the optimal model, $\mathcal{H}_{KLM}^\star=(K=3,L=3,M=2)$, is reported as having class-independent base hazard rates.

Our analysis correctly indicates that in \emph{Cohort A} only covariate 1 has a statistically significant association with hazard for the primary risk. The estimates for each of the three classes for the $N=2000$ data set, for example, are (within the estimated uncertainty) in agreement with the true values ($\ell=1$: $\beta_1^{11}=1.90 \pm 0.12$, HR=45.96, 95\% CI=[29.37,71.92], $p<10^{-7}$; $\ell=2$: $\beta_1^{21}=1.91 \pm 0.11$, HR=44.75, 95\% CI=[28.48,70.32], $p<10^{-7}$; $\ell=3$: $\beta_1^{31}=-2.16 \pm 0.14$, HR=0.013, 95\% CI=[0.008,0.023], $p<10^{-7}$). The associations for covariates 2 and 3 were correctly estimated for all of the data sets, being found to be statistically insignificant (according to $p$-value) in all of the three latent classes and the 95\% CI included the correct HR of unity in all cases. As heterogeneity is not accounted for in a standard Cox regression it should be expected that the application of such an analysis to \emph{Cohort A} would be likely to yield incorrect estimates. Indeed, Cox regression does produce a misleading interpretation of the \emph{Cohort A} data having $N=2000$, indicating that covariates 1 and 2 are both associated with an increased hazard (cov.1: HR=1.33, 95\% CI=[1.21,1.47], $p<10^{-7}$; cov. 2: HR=1.12, 95\% CI=[1.01,1.23], $p$=0.03).

The accuracy of the spline-approximated base hazard rates can be assessed both by visual inspection, for similarity between the estimated and true rates (Figure~\ref{fig:DataA_ParamEstimates}), and numerically. The time-independent base hazard rate of the third class ($\ell=3$) is reliably estimated for both data the $N=20000$ and $N=2000$ data sets. The base hazard rates of classes $\ell=1$ and $\ell=2$ are more accurately estimated for the larger data set size; the estimated base hazard rate of class $\ell=1$ begins to deviate from the actual rate for times greater than about $t=10$ for the $N=2000$ analysis, whereas deviation is not observed at times earlier than about $t=15$ for the $N=20000$ data set. Additionally, the suitability of a spline-approximation to describe the exponentially increasing base hazard rate of the second class ($\ell=2$) can be gauged by comparing the known values of the parameters, $\lambda_0$ and $\alpha$, obtained on fitting the generalised form of the exponentially increasing base hazard rate, $\lambda(t)=\lambda_0 \rme^{\alpha t}$, to the estimated base hazard rate; for the $N=20000$ data set these parameters were determined to be $\lambda_0=0.012$ and $\alpha=0.233$ and for the $N=2000$ data set they were found to be $\lambda_0=0.014$ and $\alpha=0.227$ (the true values of these parameters are $\lambda_0=0.01$ and $\alpha=0.25$).

The performance of retrospective class allocation will, of course, depend on the exact characteristics underlying a cohort and the accuracy of the optimal model representing that cohort; it would be reasonable to expect that retrospective class assignment should perform most effectively when the differences between latent classes are pronounced and that the algorithm may struggle when the latent classes do not differ significantly. The random assignment of patients into the three classes of \emph{Cohort A} would offer an accuracy of 33\% whereas 73\% of individuals were allocated correctly on applying our class assignment algorithm to the $N=20000$ data set; 72\% and 66\% of those individuals allocated to the first and second classes ($\ell=1$, $\ell=2$) actually belonged to those classes, and 84\% of individuals allocated to the third class ($\ell=3$) actually belonged to that class. Class assignment applied to the $N=2000$ data set yielded an overall accuracy of 72\% and with a similar accuracy for those assigned to the first and second class as was obtained with the $N=20000$ data set; 78\% of those allocated to the third class actually belonged to that class. It is interesting to note that for the $N=200$ data set, despite there being insufficient information for all associations and the base hazard rates to be accurately estimated, overall 70\% of individuals were correctly allocated to their true class.

\begin{figure}[t]
\centering
\includegraphics{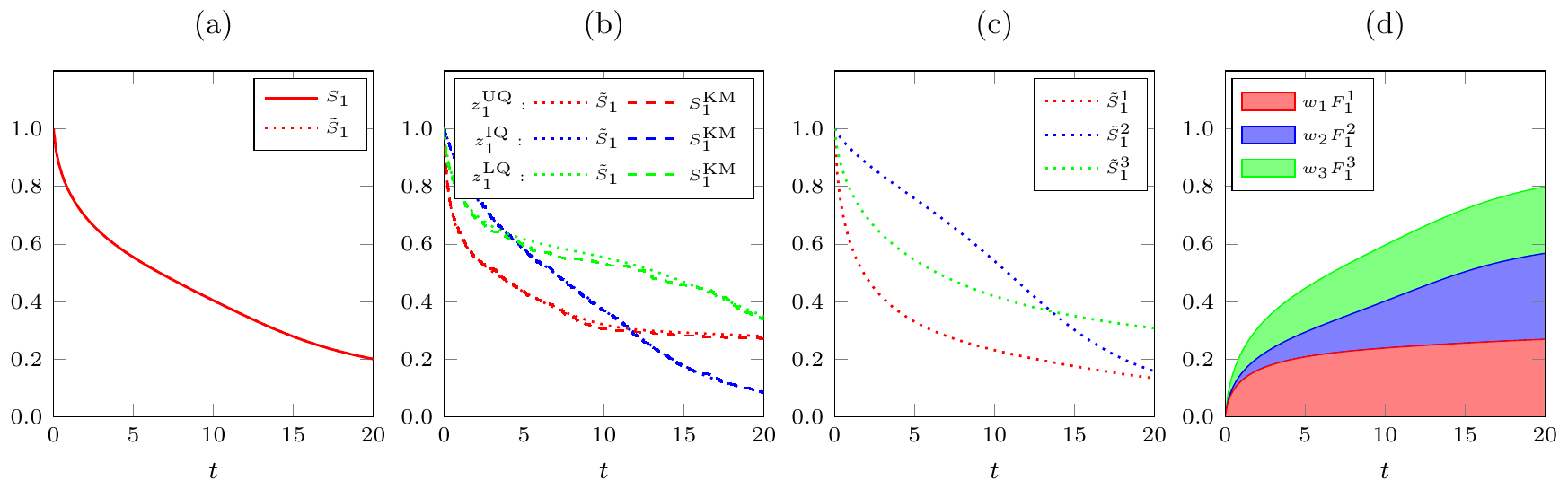}
\caption{\footnotesize{\emph{Latent class-specific survival in a heterogeneous cohort:} Survival and cumulative incidence estimates arising from the optimal parameter estimates for analysis of \emph{Cohort A} data having $N=2000$. In (a) the crude and decontaminated survival curves for the primary risk for the cohort as a whole; as there is no informative censoring in \emph{Cohort A} these curves are identical. In (b) the decontaminated survival curves, $\tilde{S}_1$, and the Kaplan-Meier estimators, $S_1^{\rm{KM}}$, for the lower quartile (LQ), upper quartile (UQ), and the inter-quartile range (IQ) of covariate 1; it is clearly apparent that proportional hazards assumption does not hold for \emph{Cohort A} as a whole. In (c) the class-specific decontaminated survival curves, $\tilde{S}_1^{\ell}$, for the three classes; the impact of the constant base hazard rate of class $\ell=1$ and the exponentially increasing base hazard rate of class $\ell=2$ is easily observed in the difference between the survival curves for the two classes. In (d) the stacked weighted class-specific cumulative incidence. }
}
\label{fig:DataA_N2000_SurvCurves}
\end{figure}

In Figure~\ref{fig:DataA_N2000_SurvCurves} class-specific survival and weighted cumulative incidence functions generated using the optimal parameter estimates for the $N=2000$ data set are shown. As there is only one real risk in \emph{Cohort A} the crude and decontaminated survival functions, $S_1$ and $\tilde{S}_1$, are identical (Figure~\ref{fig:DataA_N2000_SurvCurves}~(a)). As there is no informative censoring in \emph{Cohort A} the Kaplan-Meier estimators offer a reasonable approximation to the risk-specific survival function (Figure~\ref{fig:DataA_N2000_SurvCurves}~(b)); the crossing of the Kaplan-Meier estimators conditioned on the value of covariate 1 would suggest that there is latent heterogeneity in \emph{Cohort A} and that the application of standard Cox analysis would yield misleading estimates. The differences in the expected survival between the three classes over the duration of the trial is readily apparent on inspecting the class-specific decontaminated survival functions (Figure~\ref{fig:DataA_N2000_SurvCurves}~(c)); the significantly more rapid expiration of the first class ($\ell=1$) against the primary risk survival in comparison to that of the second class ($\ell=2$) being due to the two classes having very different base hazard rates. The weighted class-specific cumulative incidence (Figure~\ref{fig:DataA_N2000_SurvCurves}~(d)) also illustrate at which stages of the trial members of the first ($\ell=1$) and second class ($\ell=2$) succumb to the primary risk.

\subsection{Effective analysis in the presence of informative censoring}
\label{ssec:SynthData_InformativeCensoring}

The ability of our analysis to accurately model the true survival in the presence of heterogeneity-induced informative censoring is now examined. The analysis of two cohorts having identical primary risk characteristics are presented in this section; one cohort being modelled to include the effects of false protectivity and  the other to include the influence of false-aetiology.

Both \emph{Cohort B} and \emph{Cohort C} have two latent classes ($L=2$) of equal size and have personalised cause-specific hazard rates having class-dependent associations (i.e. of type $M=2$); both are subject to two real risks, the primary risk ($r=1$) and a competing risk ($r=2$), and end-of-trial censoring. The base hazard rates are time-independent but differ between the two risks. The properties of \emph{Cohort B} and \emph{Cohort C} are given in Table~\ref{tab:CohortBandCParamEstimates} alongside the Bayesian-determined optimal parameter values for data having $N=1500$ observations.

\begin{table}
\caption{{{Modelling heterogeneity-induced informative censoring:} The estimated weights, $w_\ell$, and association parameters, $\beta_{r}^{\ell \mu}$, for the two real risks $r=1$ and $r=2$, from the Bayesian-determined optimal model from analysis of the \emph{Cohort B} and \emph{Cohort C} data sets with $N=1500$. The true parameter values are shown in brackets to the right of the estimated values.}
\label{tab:CohortBandCParamEstimates}}
\vspace{3mm}
\centering
\begin{tabular}{lrrrr}
\hline
\multicolumn{5}{l}{\emph{Cohort B} $(K=1, L=2, M=2)$}\\
\multicolumn{5}{l}{Heterogeneity-induced false protectivity} \\
\hline
 &\multicolumn{2}{l}{Class, $\ell=1$} &\multicolumn{2}{l}{Class, $\ell=2$} \\
\hline
$w_{\ell}$ & $0.51 \pm 0.02$ & (0.5) & $0.49 \pm 0.02$ & (0.5) \\
\hline
$\beta_{1}^{\ell 1}$ & $1.85 \pm 0.14$ & (2.0) & $-1.97 \pm 0.10$ & (-2.0) \\
$\beta_{1}^{\ell 2}$ & $-0.05 \pm 0.10$ & (0.0) & $-0.04 \pm 0.07$ & (0.0) \\
$\beta_{1}^{\ell 3}$ & $0.15 \pm 0.10$ & (0.0) & $0.01 \pm 0.09$ & (0.0) \\
\hline
$\beta_{2}^{\ell 1}$ & $3.17 \pm 0.10$ & (3.0) & $-0.10 \pm 0.07$ & (0.0) \\
$\beta_{2}^{\ell 2}$ & $-0.07 \pm 0.08$ & (0.0) & $0.03 \pm 0.06$ & (0.0) \\
$\beta_{2}^{\ell 3}$ & $-0.06 \pm 0.07$ & (0.0) & $0.01 \pm 0.06$ & (0.0) \\
\hline
\end{tabular}
\quad
\begin{tabular}{lrrrr}
\hline
\multicolumn{5}{l}{\emph{Cohort C} ($K=1, L=2, M=2$)}\\
\multicolumn{5}{l}{Heterogeneity-induced false aetiology} \\
\hline
 &\multicolumn{2}{l}{Class, $\ell=1$} &\multicolumn{2}{l}{Class, $\ell=2$} \\
\hline
$w_{\ell}$ & $0.51 \pm 0.02$ & (0.5) & $0.49 \pm 0.02$ & (0.5) \\
\hline
$\beta_{1}^{\ell 1}$ & $2.04 \pm 0.12$ & (2.0) & $-1.94 \pm 0.10$ & (-2.0) \\
$\beta_{1}^{\ell 2}$ & $-0.06 \pm 0.07$ & (0.0) & $-0.01 \pm 0.06$ & (0.0) \\
$\beta_{1}^{\ell 3}$ & $-0.02 \pm 0.08$ & (0.0) & $0.14 \pm 0.06$ & (0.0) \\
\hline
$\beta_{2}^{\ell 1}$ & $-3.02 \pm 0.09$ & (-3.0) & $0.11 \pm 0.08$ & (0.0) \\
$\beta_{2}^{\ell 2}$ & $-0.02 \pm 0.06$ & (0.0) & $0.05 \pm 0.07$ & (0.0) \\
$\beta_{2}^{\ell 3}$ & $0.04 \pm 0.07$ & (0.0) & $-0.03 \pm 0.06$ & (0.0) \\
\hline
\end{tabular}
\end{table}

The optimal model was correctly determined for the analysis of both \emph{Cohort A} and \emph{Cohort B} data, describing heterogeneous cohorts containing two classes and with personalised hazard rates having heterogeneous associations but with class-independent  and time-independent base hazard rates. The predicted class sizes, associations, and base hazard rates for both data sets were in close agreement with their actual values, as detailed in Table~\ref{tab:CohortBandCParamEstimates}.

The Bayesian-determined estimates for both \emph{Cohort B} and \emph{Cohort C} suggest correctly that covariate 1 is strongly associated with an increased hazard for the primary risk in one class and is associated with a reduced hazard in the other class; covariates 2 and 3 were found to be statistically insignificant, according to their respective $p$-values, for both classes in both cohorts.

The Bayesian-determined estimates for the secondary risk, responsible for informative censoring, were also found to be consistent with the expected values for both cohorts. In \emph{Cohort B} covariate 1 was found to be strongly associated with an increased hazard for the secondary risk in one class only ($\ell=1$: HR=565.28, 95\% CI=[387.72,824.17], $p=10^{-7}$), and in \emph{Cohort C} covariate 1 was found to be associated with a reduced hazard for the secondary risk for one class only ($\ell=1$: HR=0.002, 95\% CI=[0.002,0.003], $p=10^{-7}$). Again, the associations for the other covariates were found to be statistically insignificant and the predicted hazard ratios suggested neither elevated or reduced hazard for the secondary risk.

It is reassuring that our analysis was able to correctly characterise the associations for the primary risk as being identical (within the estimated uncertainty of the parameter estimates) both in the presence of heterogeneity-induced false protectivity (\emph{Cohort B}) and in the presence of heterogeneity-induced false aetiology (\emph{Cohort C}). As was found in Section~\ref{ssec:SynthData_CohortSubstructure} for the analysis of heterogeneous cohort data, standard Cox regression offered misleading estimates on the analysis of \emph{Cohort B} and \emph{Cohort C} data.

\begin{figure}[t]
\centering
\includegraphics{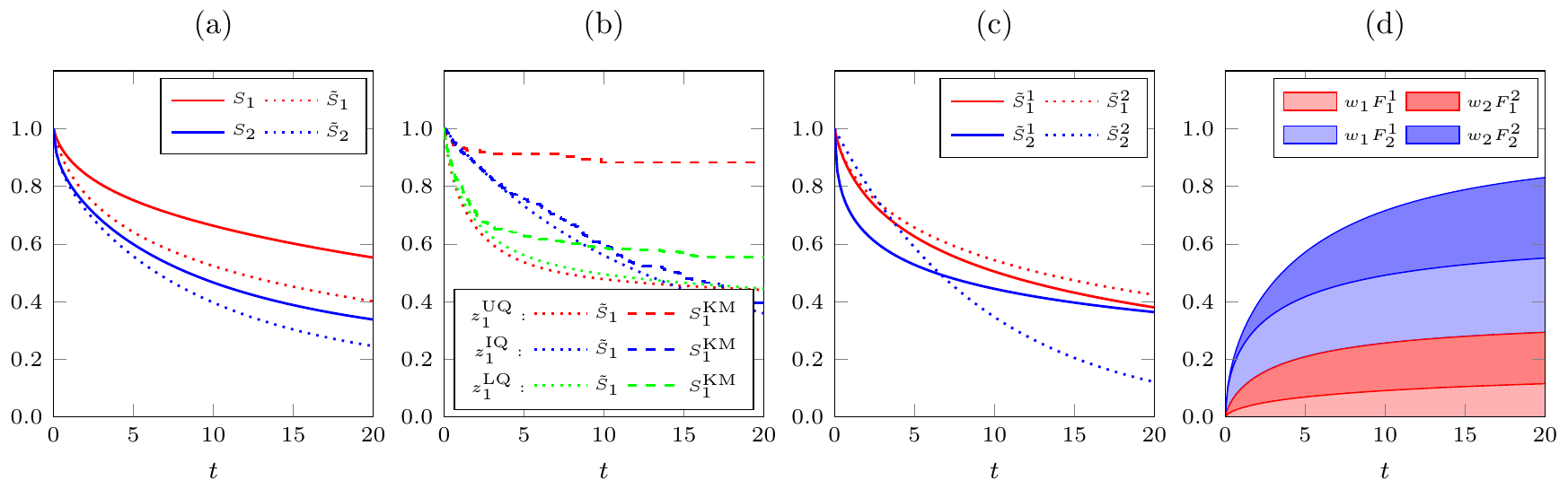}
\caption{\footnotesize{\emph{Decontaminating false protectivity effects from survival data:} Survival and cumulative incidence curves for analysis of \emph{Cohort B} (Table~\ref{tab:CohortBandCParamEstimates}) data having $N=1500$ demonstrating the effectiveness of our analysis to expose the influence of false protectivity in survival data. In (a) the crude and decontaminated risk-specific survival curves, $S_r(t)$ and $\tilde{S}_r(t)$, for the primary and secondary risks; the extent to which the secondary risk causes false protectivity is clearly apparent on observing the difference between the crude and decontaminated survival curves for the primary risk. In (b) the decontaminated survival curves, $\tilde{S}_1$, and the risk-specific Kaplan-Meier estimators, $S_1^{\rm{KM}}$, for the lower quartile (LQ), upper quartile (UQ), and the inter-quartile range (IQ) of covariate 1; it is clear that proportional hazards assumption does not hold for \emph{Cohort B} as a whole and it is striking that the risk-specific Kaplan-Meier estimator for the upper quartile of covariate 1 is extremely misleading. In (c) the class- and risk-specific decontaminated survival curves, $\tilde{S}_r^{\ell}$, for the two latent classes for both risks. In (d) the stacked weighted class-specific cumulative incidence.}}
\label{fig:DataB_ParamEstimates}
\end{figure}

The various survival estimators obtained with the optimal parameter estimates for \emph{Cohort B} are shown in Figure~\ref{fig:DataB_ParamEstimates}. False protectivity in the cohort is clearly indicated (Figure~\ref{fig:DataB_ParamEstimates}~(a)); the crude survival function for the primary risk, $S_1$, exceeds its decontaminated counterpart, $\tilde{S}_1$, consequently \emph{true} survival against the primary risk is poorer than the crude survival function suggests. In the presence of informative censoring the Kaplan-Meier estimators for the primary risk, $S_1^{\rm{KM}}$, conditioned on the value of covariate 1, are also seen to be misleading (Figure~\ref{fig:DataB_ParamEstimates}~(b)); the risk-specific Kaplan-Meier estimators follow the crude survival function (not shown) and yield extremely poor estimates for the lower and upper quartiles of covariate 1 due to risk correlations in the cohort. The Kaplan-Meier estimator, for example, indicates that individuals having an upper quartile value of covariate 1 are likely to have a relatively good survival (about 90\% surviving at the end-of-trial time) against the primary risk whereas such individuals actually have a much poorer survival (less than 50\% survival at the end-of-trial time). The class-specific survival functions and weighted class-specific cumulative incidence are also shown (Figure~\ref{fig:DataB_ParamEstimates}~(c,d)).

\begin{figure}[t]
\centering
\includegraphics{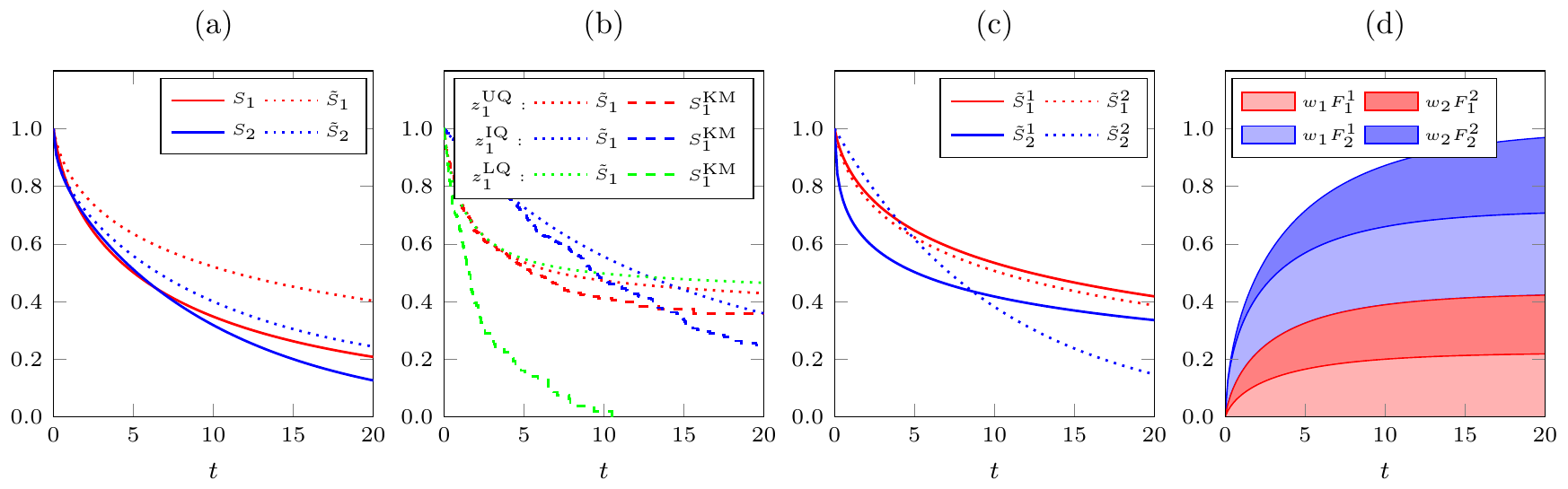}
\caption{\footnotesize{\emph{Decontaminating false aetiology effects from survival data:} Survival and cumulative incidence curves for analysis of \emph{Cohort C} (Table~\ref{tab:CohortBandCParamEstimates}) data having $N=1500$ demonstrating the effectiveness of our analysis to expose the influence of false aetiology in survival data. In (a) the crude and decontaminated risk-specific survival curves, $S_r(t)$ and $\tilde{S}_r(t)$, for the primary and secondary risks; the extent to which the secondary risk suggests false exposure is clearly apparent on observing the difference between the crude and decontaminated survival curves for the primary risk. In (b) the decontaminated survival curves, $\tilde{S}_1$, and the risk-specific Kaplan-Meier estimators, $S_1^{\rm{KM}}$, for the lower quartile (LQ), upper quartile (UQ), and the inter-quartile range (IQ) of covariate 1; it is clear that proportional hazards assumption does not hold for \emph{Cohort C} as a whole and that the risk-specific Kaplan-Meier estimator for the lower quartile of covariate 1 is extremely misleading. In (c) the class- and risk-specific decontaminated survival curves, $\tilde{S}_r^{\ell}$, for the two latent classes for both risks. In (d) the stacked weighted class-specific cumulative incidence.}
}
\label{fig:DataC_ParamEstimates}
\end{figure}

In Figure~\ref{fig:DataC_ParamEstimates} the survival estimators for \emph{Cohort C} are shown.
The influence of false aetiology effects in \emph{Cohort C} is clear on noting that the decontaminated survival function for the primary risk, $\tilde{S}_1$, exceeds the crude survival function, $S_1$, throughout the trial duration; the \emph{true} survial against the primary risk is significantly greater than the crude survival function suggests. The covariate-conditioned risk-specific Kaplan-Meier estimators prove also to be very misleading in the presence of false exposure effects in \emph{Cohort C}; the Kaplan-Meier estimator for individuals having a lower quartile value of covariate 1 suggests no survival beyond about half of the trial duration whereas the survival for such individuals is actually almost 50\% at the end-of-trial time. The class-specific survival functions and weighted class-specific cumulative incidence are also shown (Figure~\ref{fig:DataC_ParamEstimates}~(c,d)).

\section{Applications to prostate cancer data}
\label{sec:ULSAM}

Prostate cancer (PC) data are notorious for exhibiting competing risk effects \cite{Grundmark}, largely due to the fact that the disease occurs late in life when there is an increased number of non-primary events whose incidence could correlate with prostate cancer. Here we analyse data from the ULSAM cohort \cite{ULSAM} and compare the outcomes of Cox's proportional hazards regression \cite{Cox} and our present method. The ULSAM cohort has $N=2047$ individuals of which 208 reported PC as the first event as described previously \cite{ULSAMpaper} and we have included five relevant covariates of the ULSAM data \cite{ULSAM}.\footnote{Since this study does not seek to answer a pre-specified clinical hypothesis, the actual choice made for the set of covariates to be included is not critical.}

Smoking is suggested to have a weak protective effect against PC risk (HR=0.85, 95\% CI=[0.65,1.11], $p$=0.23) in the ULSAM cohort according to Cox's proportional hazards analysis. However, as the Kaplan-Meier estimator for PC risk conditioned on the smoking covariate (see Figure~\ref{fig:synthetic_illustration}) does not meet the proportional hazards assumption, any conclusions drawn from such an analysis (on the ULSAM cohort data as a whole, at least) may be invalid. The non-proportionality of the smoking covariate-conditioned Kaplan-Meier estimator for PC risk could be due to heterogeneity in the ULSAM cohort.

\begin{figure}[t]
\centering
\includegraphics{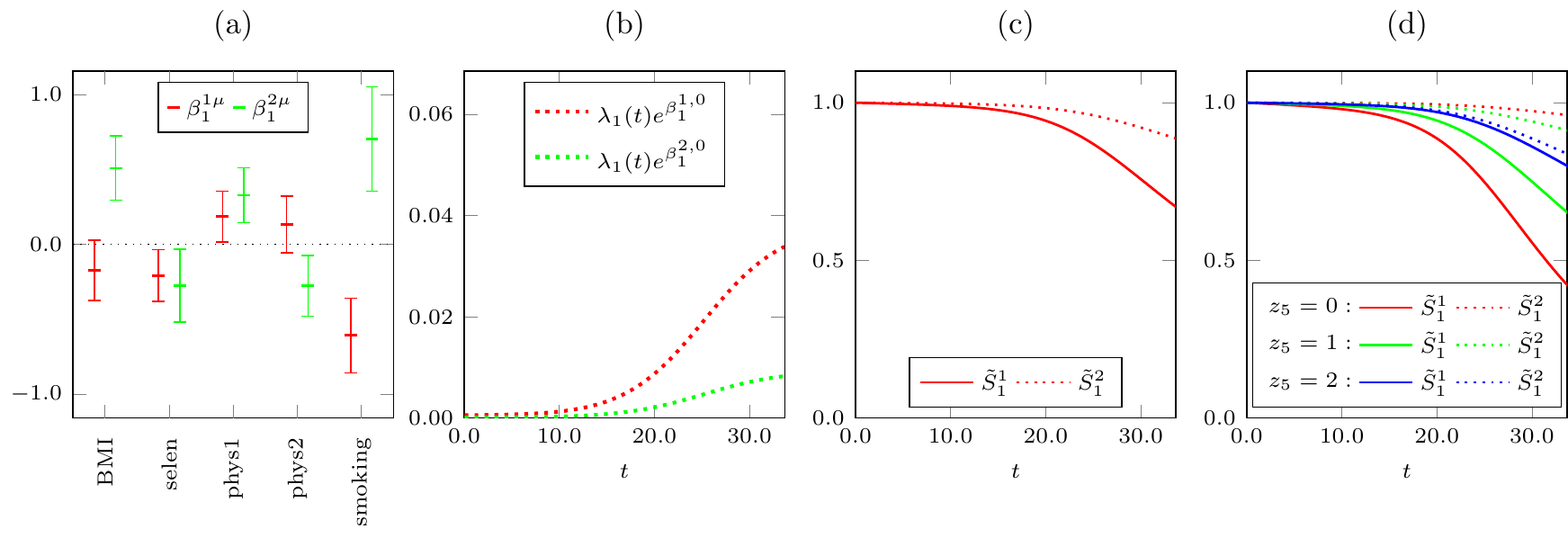}
\caption{\footnotesize{\emph{Bayesian-determined heterogeneous characterisation of the ULSAM cohort:} In (a) and (b) the associations and frailty weighted-base hazard rates for PC risk, for the two latent classes of the optimal model $\mathcal{H}_{KLM}^\star=(K=3,L=2,M=2)$ obtained on analysis of the ULSAM cohort data. The included covariates are: body mass index (real-valued), serum Selenium level (\emph{selen}, integer valued), leisure time physical activity (\emph{phys1}, discrete levels 0/1/2), work physical activity (\emph{phys2}, discrete levels 0/1/2), and smoking status (\emph{smoking}, discrete levels 0/1/2). The estimated weighting of the classes for this model is $w_{1}=0.32 \pm 0.08$, $w_{2}=0.68 \pm 0.08$. The retrospective weighting of the classes for this model is $f_{1}=0.16$ (332 of 2047 patients), $f_{2}=0.84$ (1715 of 2047 patients). In (c) the class-and risk-specific survival functions; for PC risk it is clear that the survival of members of the frailer class ($\ell=1$) is poorer than those of the healthier class ($\ell=2$), significantly and increasingly so for times beyond about 20~years. In (d) the class-and risk-specific survival function conditioned on the three values of the smoking covariate ($z_5=0$: non-smoker, $z_5=1$: ex-smoker, $z_5=2$: smoker); it is clear that for the majority of the cohort ($\ell=2$) smoking is associated with poorer survival outcomes but that for those members of the frailer class ($\ell=1$) smokers and ex-smokers have greater survival than non-smokers.}
}
\label{fig:UlsamEstimatesAndSurvCurves}
\end{figure}

The optimal characterisation of the ULSAM data, as determined by Bayesian model selection, suggests that the cohort should be viewed as consisting of two distinct classes: one class $\ell=1$ with relatively frail individuals (in terms of both primary and secondary risk) which contains about 16\% of the cohort according to retrospective class allocation, and another class $\ell=2$ with rather healthy individuals which contains the remainder of the cohort. The estimated association parameters, base hazard rates, and class-specific survival functions for PC are shown in Fig.~\ref{fig:UlsamEstimatesAndSurvCurves} for the most probable model ($\mathcal{H}_{KLM}^\star=(K=3,L=2,M=2)$). The decontaminated survival function for PC was found to be marginally less than the crude survival function for follow-up times exceeding about 20~years; this is indicative of false protectivity effects for PC risk in the ULSAM cohort data.

In the Bayesian-determined two-class description of the ULSAM cohort, BMI and smoking are recognised as serious PC risk factors for those individuals in the healthier class. In the healthier class ($\ell=2$) smoking is associated with elevated risk of PC (HR=4.08, 95\% CI=[1.09,15.29], $p$=0.04), as is having a higher BMI (HR=2.77, 95\% CI=[1.20,6.36], $p$=0.02). Conversely, in the frailer class ($\ell=1$) smoking is associated with a decreased risk of PC (HR=0.30, 95\% CI=[0.12,0.76], $p$=0.01). In the frail class the regression coefficients are weaker than those of the stronger class, and one expects the negative coefficients for e.g. BMI and smoking to reflect reverse causality: within this group, having a higher BMI and still being {\em able} to smoke may well be an indicator of {\em relatively} good health.

Our explanation of the ULSAM data is not necessarily the final one. There are alternative ways to do the regression, e.g. by combining all non-primary risks (including the end-of-trial risk). The conclusion to be drawn is that the new two-class explanation of the ULSAM data is both probabilistically and intuitively more plausible.

\section{Applications to breast cancer data}
\label{sec:Amoris}

In this section the results of the application of our analysis to data from the Swedish Apolipoprotein Mortality Risk Study (AMORIS)  e.g. \cite{Holme_2010,Holme_2008} are summarised. The analysed data set is that for which a competing risk analysis is presented in \cite{WahyuWulan_2015}, and contained data for $N=1798$ women from the AMORIS population for whom baseline serum glucose, triglyceride, and total cholesterol measurements were available within three months to three years prior to breast cancer diagnosis. The time-to-event describes the duration between diagnosis and breast cancer death (BC), cardiovascular disease death (CV), death from other causes, or departure from or the end of the study (censoring) and age, fasting status, and socio-economic status data were also included as covariates.

\begin{figure}[thb]
\centering
\includegraphics{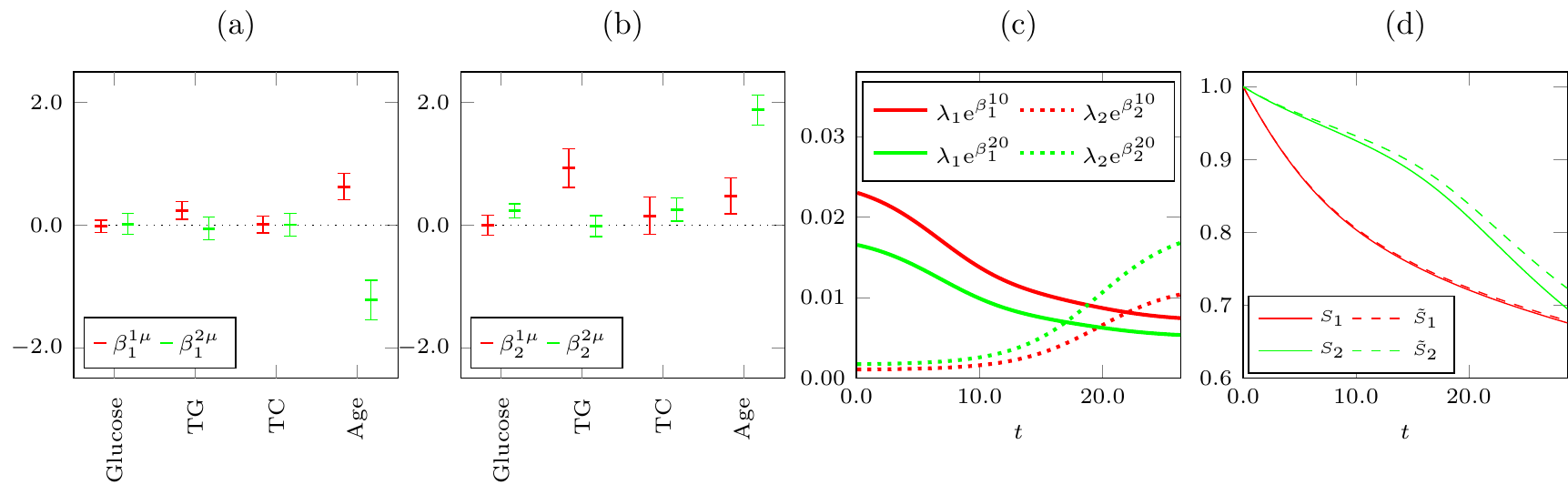}
\caption{\footnotesize{\emph{Bayesian-determined heterogeneous characterisation of a cohort of women from the AMORIS population diagnosed with breast cancer:} Parameter estimates for the Bayesian-determined optimal model, $\mathcal{H}^\star_{KLM}=(K=3,L=3,M=2)$, obtained on analysis of data for $N=1798$ women from the AMORIS population diagnosed with breast cancer; the estimated retrospective weighting of the classes for this model is $f_{1}=0.66$ (1189 of 1798 patients), $f_{2}=0.32$ (574 of 1798 patients), $f_{3}=0.02$ (35 of 1798 patients). In (a) and (b) the estimated hazard associations, $\beta_r^{\ell\mu}$, with serum glucose, triglyceride (TG), total cholesterol (TC) and age, for BC death ($r=1$) and CV death ($r=2$) respectively for the dominant ($\ell=1$: red) and next largest ($\ell=2$: green) latent classes. In (c) the base hazard rates for BC and CV death for the two dominant latent classes; note that the base hazard rate for CV death exceeds that for BC death for follow-up times greater than about 15~years ($\ell=1$: red, $\ell=2$: green). In (d) the crude (solid lines) and decontaminated (dashed lines) cause-specific survival against BC (red lines) and CV death (green lines); observe that the crude survival against CV death, $S_{2}$, is poorer than its decontaminated counterpart, $\tilde{S}_{2}$.}
}
\label{fig:AmorisEstimates}
\end{figure}

Our analysis suggests that the cohort of women from the AMORIS population diagnosed with breast cancer is best described by three latent classes which share the same general trend of having base hazard rates which decrease with time for BC death risk but increase with time for CV death risk. The largest class ($\ell=1$) contains about 66\% of the cohort according to retrospective class allocation and has a poorer survival against BC death but a greater survival against CV death than does the second class ($\ell=2$), which accounts for 32\% of the cohort. The smallest latent class ($\ell=3$) is almost wholly comprised of relatively younger individuals for whom death from other causes was reported; it is likely that the inclusion of this third class, containing only 35 of the 1798 individuals according to retrospective class assignment, enables more effective characterisation of the other two classes as it allows the effects of two differing groups within those reported as having death from other causes to be distinguished. As BC and CV death shall be the focus of the remainder of this section, parameter estimates and survival curves pertaining to the small third latent class ($\ell=3$) are omitted in the interest of readability. The estimated associations, base hazard rates, and decontaminated class-specific survival curves are shown in Figure~\ref{fig:AmorisEstimates} for the optimal characterisation of this cohort.

The largest class ($\ell=1$) has the greatest hazard for BC death, having a base hazard rate which exceeds that of the second class ($\ell=2$) over the entire interval (Figure~\ref{fig:AmorisEstimates}~(c)); the opposite is true for the hazard for CV death, with the second class ($\ell=2$) having the greatest hazard for CV death. The base hazard rate for BC death is most significant at diagnosis ($t=0$) and diminishes with time whereas the risk of CV death increases with time. The base hazard for CV death exceeds the risk of BC death after about 20~years from diagnosis for those in the first class ($\ell=1$) and after about 16~years for those in the second class ($\ell=2$); this is generally consistent with previous findings regarding the importance of considering co-morbidities for women with breast cancer \cite{Patnaik_2011}. Accordingly, survival against BC death is greater for the less frail second class ($\ell=2$) than for the first class ($\ell=1$) and survival against CV death is greater in the first class ($\ell=1$) than in the second class ($\ell=2$).

In the relatively frail first class ($\ell=1$), triglyceride levels were found to be associated with an increased hazard for both CV death (CV, $\ell=1$: HR=6.44, 95\% CI=[1.86,22.29], $p=0.003$) and for BC death (BC, $\ell=1$: HR=1.62, 95\% CI=[0.92,2.84], $p=0.09$). Standard Cox analysis suggests a weaker association with an increased hazard against CV death (CV, Cox: HR=1.61, 95\% CI=[1.17,2.21], $p=0.003$) than is suggested by our analysis, a probable consequence of those members of the cohort for which triglyceride levels are not associated with increased hazard ($\ell=2$) diluting the effects of those for which triglyceride levels are associated with increased hazard ($\ell=1$); the same is also true for BC death, for which standard Cox analysis indicates that triglyceride levels are associated with a modestly increased hazard for BC death (BC, Cox: HR=1.22, 95\% CI [0.98,1.52], $p$=0.076). Serum glucose was found to be associated with an increased hazard for CV death (CV, $\ell=2$: HR=1.59, 95\% CI=[1.01,2.49], $p=0.04$) for those in the relatively less frail second class ($\ell=2$), as was total cholesterol (CV, $\ell=2$: HR=1.65, 95\% CI=[0.79,3.51], $p=0.18$).

\begin{figure}[t]
\centering
\includegraphics{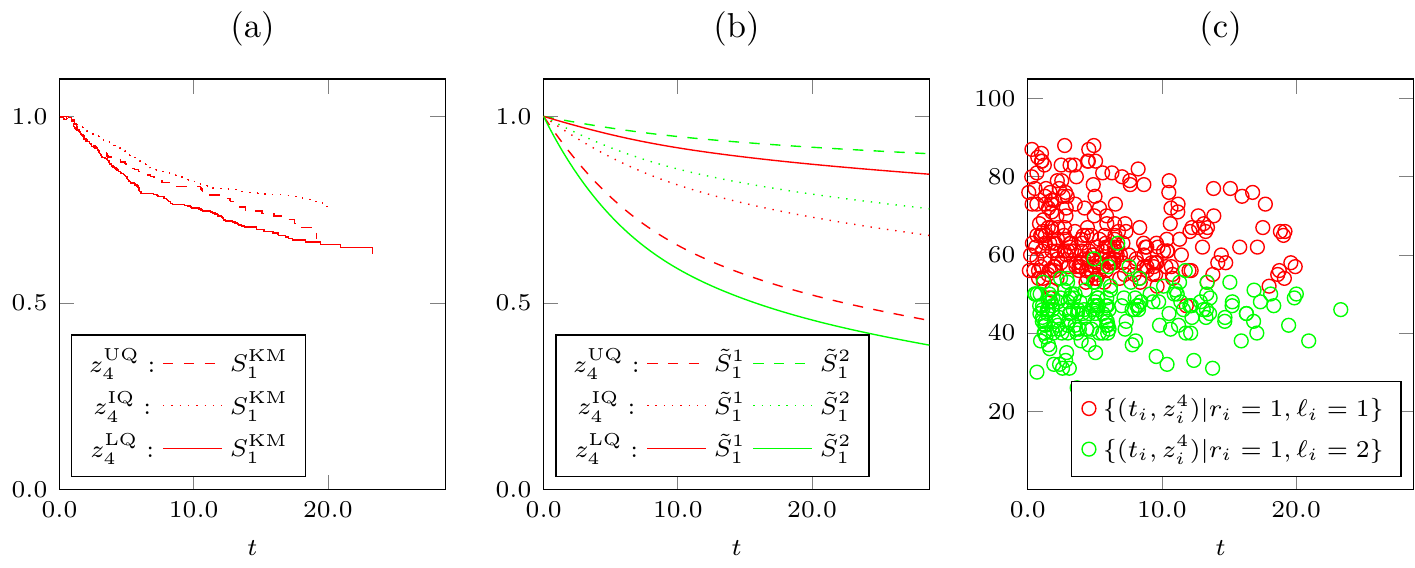}
\caption{\footnotesize{\emph{Age and survival in women from the AMORIS population diagnosed with breast cancer:} In (a) the age-conditioned Kaplan-Meier estimator for BC death; observe that survival of both the upper and lower quartiles of age the proportional hazards is poorer than that for the inter-quartile range for age (i.e. the proportional hazards assumption is not met) and is suggestive of heterogeneity in the cohort. In (b) the class-specific decontaminated survival curves, $\tilde{S}_r^\ell$ conditioned on the age covariate ($z_4$), for BC death ($r=1$). The substantial difference in survival against BC death between those in the lower quartile (LQ) and upper quartile (UQ) for age is indicative of the strength of the association of age with hazard for BC death (see Figure~\ref{fig:AmorisEstimates}). Observe that in the more frail first class ($\ell=1$) survival against BC death decreases with increasing age while the opposite is true of the second class ($\ell=2$); however, survival of the second class ($\ell=2$) is actually poorer for younger individuals than it is for older from the more frail first class ($\ell=1$). In (c) the time-to-event and age for each individual having succumbed to BC death is shown, indicating that there is a clear stratification, according to age, of the latent classes to which individuals are retrospectively assigned. Almost all of those individuals younger than about 50~years old are assigned to the second latent class ($\ell=2$) whereas those older than about 50~years are assigned to the first class ($\ell=1$).}
}
\label{fig:AmorisClassesAndSurvCurves}
\end{figure}

In contrast to standard Cox analysis, which suggests that age is associated with a reduced hazard for BC death (BC, Cox: HR=0.77, 95\% CI=[0.59,1.00], $p$=0.05) for the cohort as a whole, our analysis indicates that for the majority of the cohort ($\ell=1$) age is associated with an increased hazard for BC death (BC, $\ell=1$: HR=3.49, 95\% CI=[1.51,8.10], $p$=004) and is associated with a reduced hazard for BC death (BC, $\ell=2$: HR=0.09, 95\% CI=[0.02,0.30], $p\approx 10^{-4}$) for about only one third of the cohort ($\ell=2$). The age of an individual was also found to be associated with their hazard for CV death, most significantly being strongly associated with an increased hazard for CV death (CV, $\ell=2$: HR=42.93, 95\% CI=[16.43,112.21], $p< 10^{-7}$) in the less frail second class ($\ell=2$). The age-conditioned risk-specific Kaplan-Meier estimator and class-specific decontaminated survival for BC death are shown in Figure~\ref{fig:AmorisClassesAndSurvCurves}, along with the age, time-to-event, and retrospectively assigned class, for those individuals recorded as succumbing to BC death. The stratification by age, between about 50 and 55~years, of those individuals succumbing to BC death and retrospectively allocated to the relatively frail class ($\ell=1$) and those allocated to the second class ($\ell=2$), suggest that menopausal status may be informative as to expected survival against BC death, as shown in Figure~\ref{fig:AmorisClassesAndSurvCurves}(c).

Survival against BC death for the more frail first class ($\ell=1$) is greatest for the younger members of the cohort, with over 80\% survival for those in the LQ for age in the cohort as opposed to less than 50\% survival for those in the UQ for age. In the second class ($\ell=2$) survival against BC death is markedly poorer for younger members of the cohort as a consequence of age being strongly associated with reduced hazard for BC death for this class.

\section{Discussion}
\label{sec:Discussion}
 
In this study we have introduced a fully Bayesian approach for the determination of the probabilistically optimal characterisation of a complex cohort. The introduction of the concept of heterogeneity-induced informative censoring, where risks are independent only at the level of individuals, has enabled us to identify the `decontaminated' hazard rates and survival functions.

The software package, \emph{ALPACA}, which implements the formalism introduced herein provides practical tools for survival analysis which can be applied to heterogeneous cohorts and in the presence of heterogeneity-induced informative censoring. This implementation is able to find the most probable characterisation of a cohort through the combination of i) a latent class model which captures, for \emph{all} risks and for \emph{each} class, the base hazard rate, covariate association(s) and relative frailty, and ii) Bayesian model selection to determine the optimal number of classes, the optimal parametrisation of each base hazard rate, and the extent of heterogeneity in the cohort. Once the optimal description of the cohort has been determined, crude and decontaminated cause-specific survival curves can be compared to gauge the extent of any informative censoring, and differences between class-specific survival curves can be examined. Exploration of correlations between covariate values and retrospectively assigned class membership can offer an extra insight into a cohort and may aid the search for new informative biomarkers. An additional advantage of our approach is that through the incorporation of Bayesian model selection to determine the optimal description of a cohort, our survival analysis can be applied with little interaction or effort required from the analyst prior to inspection and assessment of the frailties, covariate association(s), and base hazard rates for the optimal model.

Applied to synthetic data, our analysis was shown to effectively characterise heterogeneous cohorts, successfully remove heterogeneity-induced false protectivity and false aetiology effects, and even discriminate between two classes differing only in their base hazard rates. Retrospective class allocation was demonstrated to have an impressive accuracy even for survival data modelling a cohort containing 200 individuals.

On application to real survival data from the ULSAM cohort, with prostate cancer as the primary risk, our analysis leads to plausible alternative explanations for previous counter-intuitive inferences (such as a weak protective effect on PC of smoking), in terms of distinct sub-groups of patients with distinct risk factors and overall frailties. In the ULSAM cohort, it was also shown that the men's metabolic status introduce competing risk problems, although using traditional methods, several different analyses had to be done to reveal the underlying risk pattern \cite{Grundmark}, which with our proposed method could be done coherently in one analysis.

Applied to survival data for women diagnosed with breast cancer from the Swedish Apolipoprotein Mortality Risk Study (AMORIS), our analysis suggests that for breast cancer death risk the base hazard rate decreases with time but increases with time for cardiovascular death risk, and the age-class membership correlations may suggest differences in association patterns and survival against breast cancer between pre- and post-menopausal women.

\appendix

\section{Latent class model: additional details and identities}
\label{app:TheoryDetails}

\subsection{Connection between cohort level and individual level cause-specific hazard rates}
\label{app:link_for_rates}

\begin{sloppypar}
The relationship between the cohort level covariate-conditioned cause-specific hazard rate, $h_r(t|\bz)$, and the hazard rates of the individual members of the cohort, $h^i_r(t)$, is obtained by substitution of $\Prob(t_0,\ldots,t_R|\bz)=\sum_{i,~\bz_i=\bz}\Prob_i(t_0,\ldots,t_R)/\sum_{i,~\bz_i=\bz}1$ into (\ref{eq:covariates_h}), to give,
\end{sloppypar}

\begin{eqnarray}
h_r(t|\bz)S(t|\bz)
=
\frac{\sum_{i,~\bz_i=\bz}\int_0^\infty\!\!\!\!\ldots\!\int_0^\infty\!\dt_0\ldots \dt_R~\Prob_i(t_0,\ldots,t_R)\delta(t-t_r)\prod_{r^\prime\neq r}^R\theta(t_{r^\prime}-t)}{\sum_{i,~\bz_i=\bz}1}
=
\frac{\sum_{i,~\bz_i=\bz} S_i(t)h_r^i(t)}{\sum_{i,~\bz_i=\bz}1}.\nonumber
\end{eqnarray}

\begin{sloppypar}
\noindent Insertion of the identity $S(t|\bz)=\sum_{i,~\bz_i=\bz} S_i(t)/\sum_{i,~\bz_i=\bz}1$ and the individualised survival function, $S_i(t)=\rme^{-\sum_{r} \int_0^t\ds~h^i_r(s)}$, into the above leads to the relationship expressed by (\ref{eq:link2}).
\end{sloppypar}

\subsection{Equivalence of crude and decontaminated survival in the absence of competing risks}
\label{app:CrudeDeconNoRiskCorrelations}

In the case that a cohort is exposed to only one risk the crude and decontaminated survival, $S_r(t)$ and $\tilde{S}_r(t)$ respectively (\ref{eq:trueStilde}), are equivalent, as shown below.

As only one risk is present (i.e. $R=1$), the crude and decontaminated cohort-level hazard rates, $h_1(t|\bz)$ and $\tilde{h}_1(t|\bz)$ respectively (\ref{eq:truehtilde}), are identical. As the crude and decontaminated survival are initially equal, $S_r(t=0)=\tilde{S}_r(t=0)=1$, in order to prove that in the absence of \emph{any} competing risks are the same at any time $t$ it is sufficient to show that their time derivatives are also equal, as follows,

\begin{eqnarray}
\frac{\rmd}{\rmd t}\left[ \ln S_1(t|\bz) - \ln \tilde{S}_1(t|\bz)\right] =
-h_1(t|\bz) + \frac{\sum_{i,\bz_i=\bz} h_1^i(t) \rme^{\int_0^t \rmd s h_1^i(s)}}{\sum_{i,\bz_i=\bz} \rme^{\int_0^t \rmd s h_1^i(s)}} =
-h_1(t|\bz) + h_1(t|\bz) =
0. \nonumber
\end{eqnarray}

\subsection{Bayesian retrospective class assignment}
\label{app:ClassAssignment}

Bayesian arguments allow us to calculate class membership probabilities {\em retrospectively}\footnote{As our latent classes are defined in terms of the relation between covariates and risk, class membership for individuals cannot be predicted on the basis of covariate information alone.} for any individual for whom we have their covariates $\bz$ and survival information $(t,r)$. The probability of an individual belonging to class $\ell$, conditioned on their covariates and survival information, follows from (\ref{eq:RiskTimeProb}), and is given by,

\begin{eqnarray}
P(t,r|\bz,\ell)= 
\rme^{-{\Lambda}_0(t)} ~{\lambda}_r^{\ell}(t) \rme^{{\bbeta}^\ell_r\cdot \bz-\sum_{r^\prime=1}^R \exp({\bbeta}^\ell_{r^\prime}\cdot \bz){\Lambda}_{r^\prime}^{\ell}(t)}.
\nonumber
\end{eqnarray}

\noindent Given that $P(t,r,\ell|\bz)=P(t,r|\bz,\ell)w_\ell$ and $P(t,r|\bz)=\sum_{\ell^\prime=1}^L P(t,r|\bz,\ell^\prime)w_{\ell^\prime}$, it follows that the probability of an individual belonging to class $\ell$ is given by,

\begin{eqnarray}
P(\ell|t,r,\bz)
=  \frac{w_\ell P(t,r|\bz,\ell)}{\sum_{\ell^\prime=1}^L w_{\ell^\prime} P(t,r|\bz,\ell^\prime)}.
\end{eqnarray}

\noindent Substitution of $P(t,r|\bz,\ell)$ into the above expression leads to the retrospective class membership probability, as is given for the fully heterogeneous model variant ($M=3$) in (\ref{eq:find_class}).

\subsection{Prior distributions for the latent class model parameters}
\label{app:ModelPriors}

In seeking the optimal characterisation of a cohort using Bayesian inference (Section~\ref{ssec:BayesianInference}) it is necessary that a prior distribution, $p(\btheta_{KLM})$, on the latent class model parameters, $\btheta_{KLM}$, be defined. This requires that suitable prior distributions be chosen to encode any available information regarding the weights, frailties, associations, and those parameters used in the base hazard rate approximation.

The maximum entropy prior on the $L$ weight parameters has been used, following from the constraint $\sum_{\ell} w_{\ell}=1$, and is given by $p(w_1,\dots,w_L)=1/Z_L$, where $Z_L=\int_0^1 \!\dw_1\dots\int_0^1 \!\dw_L~\delta(\sum_\ell w_\ell - 1)=1/(L-1)!$. Unit-variance zero-average Gaussian priors were chosen for the frailty and association parameters; this is justified by our decision to pre-process the data such that all covariate distributions are normalised (by linear rescaling) to zero average and unit variance over the cohort.

\subsection{Model evidence determination using a Gaussian approximation}
\label{app:ModelEvidenceGaussianApprox}

\begin{sloppypar}
To find the optimal characterisation of a cohort using Bayesian model selection requires that the model supported by the greatest ``evidence" be determined. The posterior distribution can be written as $P(\btheta|D)=Z_{KLM}^{-1} \rme^{-S(\btheta,D)}$, where $S(\btheta,D)=-\ln[P(\btheta|\mathcal{H}_{KLM}) P(D|\btheta,\mathcal{H}_{KLM})]$, and the model evidence is proportional to $Z_{KLM}$, as given by
\end{sloppypar}

\begin{eqnarray}
Z_{KLM} =
\int \!\!\rmd \btheta ~P(\btheta|\mathcal{H}_{KLM}) P(D|\btheta,\mathcal{H}_{KLM})
\label{eq:AppZ}.
\end{eqnarray}

In our analysis, determination of the evidence (the volume of the posterior) is achieved via a Gaussian approximation to the posterior by making a Taylor expansion around its maximum, as described in e.g. \cite{MacKay}. The posterior is approximated by $P(\btheta|D,\mathcal{H}_{KLM}) = Z_{KLM}^{-1} \exp[-S(\btheta^\star)-(\btheta-\btheta^\star).A(\btheta-\btheta^\star)/2]$, where $\btheta^\star$ is the location of its maximum and $A$ is the Hessian matrix. The Gaussian-approximation for (\ref{eq:AppZ}), $Z_{KLM}^\star$, is obtained by integrating over the normalised approximated posterior distribution and recalling the standard form for a Gaussian integral, and is given by,

\begin{eqnarray}
Z_{KLM}^\star = \rme^{-S(\btheta^\star)} (2\pi)^{Y/2} (\det A)^{-1/2},
\end{eqnarray}

\noindent where $Y$ is the dimensionality of $\btheta$. In this work, the matrix $A$ is estimated numerically by investigating the curvature of the posterior distribution around its maximum.

\section{Generation of synthetic time-to-event data}
\label{app:SynthData}

Synthetic data having risks that are independent at the level of individuals and with individual cause-specific hazard rates of the form given in Table~\ref{tab:ModelVariants} was generated as described below.

The latent class, $\ell\in 1,\dots,L$, to which each individual, $i=1,\dots,N$, belongs is set at the time of generation. Covariate values for each individual, $z_i^\mu\in\mathcal{N}(0,1)$, were generated independently from a normal distribution having zero average and unit variance. A latent event time, $t_i^r$, was generated for each individual and for each risk, $r=1\ldots R$, according to $t_i^r(u)=\Lambda_r^{\ell,\rm{inv}}(\rme^{-\bbeta_r^\ell.\bz_i} \log(u))$ where $\Lambda_r^{\ell,\rm{inv}}(v)$ is the inverse of $\Lambda_r^\ell(t)=\int_0^t\!\ds ~\lambda_r^{\ell}(s)$ and $u\in[0,1]$ is a uniformly distributed random variable. The survival data for individual $i$ is that for which the latent event time is smallest $t_i={\rm min}_{r\in\{1,\ldots,R\}} t_{i}^{r}$. Should $t_i$ exceed the trial duration then individual $i$ assumes the end-of-trial censoring event and trail duration.

\end{document}